\documentclass[referee]{gji}

\usepackage{graphics}

\title[Mono Lake or Laschamp geomagnetic event]
  {Mono Lake or Laschamp geomagnetic event recorded
   from lava flows in Amsterdam Island (southeastern Indian Ocean).}

\author[Carvallo et al.]
  {Claire Carvallo${}^1$\thanks{Now at Geophysics, Department of Physics, University of Toronto, Canada}, Pierre Camps${}^1$\thanks{Corresponding author. Tel: +33 467 14 39 38; Email: camps@dstu.univ-montp2.fr}, Gilles Ruffet${}^2$, Bernard Henry${}^3$ \& T. Poidras${}^1$\\
${}^1$Laboratoire de Tectonophysique, CNRS and ISTEEM, Universit\'e Montpellier 2, Montpellier, France\\
${}^2$CNRS and G\'eosciences Rennes,  Universit\'e de Rennes 1, Rennes, France\\
${}^3$Laboratoire G\'eomagn\'etisme et Pal\'eomagn\'etisme, CNRS and IPGP, Paris, France.
  }

\date{under press in {\it Geophysical Journal International}\\
      April 17th, 2003}
\pagerange{\pageref{firstpage}--\pageref{lastpage}}
\pubyear{2003}

\begin{document}
\label{firstpage}

\maketitle

\begin{summary}
  
  We report a survey carried out on basalt flows from Amsterdam Island
  (Southeastern Indian Ocean) in order to check the presence of
  intermediate directions interpreted to belong to a geomagnetic field
  excursion within the Brunhes epoch \cite{Watkins73}, completing this
  paleomagnetic record with paleointensity determinations and
  radiometric dating.  Because the paleomagnetic sampling was done in
  few hours during the resupply of the French scientific base Martin
  du Viviers by the Marion Dufresne vessel, we could collected only 29
  samples from 4 lava flows.  The directional results corroborate the
  findings by Watkins \& Nougier~\shortcite{Watkins73} : normal
  polarity is found for two units and an intermediate direction, with
  associated Virtual Geomagnetic Poles (VGPs) close to the equator,
  for the other two units.  A notable result is that these volcanic
  rocks are well suited for absolute paleointensity determinations.
  Fifty percent of the samples yields reliable intensity values with
  high quality factors.  An original element of this study is that we
  made use of the thermomagnetic criterion PTRM-tail test of
  Shcherbakova et al.~\shortcite{Shcherbakova00} to help in the
  interpretation of the paleointensity measurements.  Doing thus, only
  the high temperature intervals, beyond 400\degr C, were retained to
  obtain the most reliable estimate of the strength of the ancient
  magnetic field.  However, not applying the PTRM-tail test does not
  change the flow-mean values significantly because the samples we
  selected by conventional criteria for estimating the paleointensity
  carry only a small proportion of their remanence below 400\degr C.
  The normal units yield Virtual Dipole Moments (VDM) of 6.2 and 7.7
  ($\mathrm{10^{22} Am^2}$) and the excursional units yield values of
  3.7 and 3.4 ($\mathrm{10^{22} Am^2}$).  These results are quite
  consistent with the other Thellier determinations from Brunhes
  excursion records, all characterized by a decrease of the VDM as VGP
  latitude decreases.  \chemical{^{40}Ar/^{39}Ar} isotopic age
  determinations provide an estimate of 26$\pm$15 Kyr and 18$\pm$9 Kyr
  for the transitional lava flows, which could correspond to the Mono
  Lake excursion.  However, the large error bars associated with these
  ages do not exclude the hypothesis that this event is the Laschamp.

\end{summary}


\section{Introduction}

Just over twenty years ago Hoffman~\shortcite{Hoffman81} concluded
from the characteristics of the Earth's magnetic field observed during
geomagnetic excursions that at least some excursions are aborted
reversals.  This implies that paleomagnetic field variations derived
from such records can be interpreted as transitional fields.
Recently, Gubbins~\shortcite{Gubbins99} proposed a physical
explanation to this hypothesis.  According to Gubbins excursions may
occur when two reversals of opposite sense follow one another in the
liquid outer core, which has timescales of about 500 yr, within such a
short time interval that the field does not have enough time to
reverse its polarity in the solid inner core, which occurs by
diffusion with typical timescales of 3 Kyr.  In other words, he
considers that the longer dynamical timescales of the solid inner core
delay full reversals which are not complete until the magnetic field
reverses its polarity throughout the whole Earth's core.  This
physical distinction between successful and unsuccessful reversals
agrees with both the duration of excursions, estimated at a few
thousands years, and with their number of about ten observed within
the Brunhes period (from 780 Kyr to present time) \cite{Langereis97}.
To validate such models one must know the exact duration of excursions
and the frequency of their occurrences between two consecutive full
reversals.  However even for the Brunhes period, we are not yet able
to precisely describe the excursion succession and their individual
characteristics.  Some excursions are not clearly established, the
global occurrence of some others is questioned~\cite{Langereis97}.
For example, the Mono Lake excursion ($\approx$ 28 ka) is observed
only in sediments from the western North
America~\cite{Liddicoat79,Levi89,Liddicoat92,Liddicoat96} and from
Arctic Sea~\cite{Nowaczyk97} and thus may only reflect a strong
regional secular variation feature.  To date only 6 excursions within
the Brunhes period seem well established, accurately dated and appear
to occur globally.  This is the case for Laschamp (40-45 Ka), Blake
(110-120 Ka), Jamaica/Pringle Falls (218 $\pm$10 ka), Calabrian Ridge
1 (315-325 ka) and 2 (515-525 ka) and Emperor/Big Lost (560-570 ka)
(see e.g. Langereis et al. (1997) and reference therein for a review).
Nevertheless it is not yet completely ascertained whether global
excursions are synchronous all over the globe or start at some
locations before spreading over the entire core surface.  The answer
to this question is of importance for both geodynamo models and
correlations of high resolution sedimentary sequences.  Furthermore,
the available data set is somewhat incomplete: most of the excursions
have been mainly described from recordings in sediments which are less
reliable than volcanic rocks in term of paleodirection, particularly
when the field intensity is low, which is a common characteristic of
the Brunhes excursions~\cite{Guyodo99}.  It is however not surprising
that the volcanic recordings are rare, since due to the sporadic
character of the volcanic extrusion rate and the short duration
assumed for excursions, the probability that a lava flow occurs during
an excursion is very small.  As a consequence, few studies have been
carried out on absolute paleointensities during Brunhes excursions
\cite{Roperch88,Chauvin89,Levi90,Quidelleur99,Schnepp94,Zhu00}.

In this context, we decided to reexamine a geomagnetic excursion
supposed to belong to the Brunhes period and previously described by
Watkins \& Nougier~\shortcite{Watkins73} in lava flows from Amsterdam
Island (southeastern Indian Ocean).  This event has never been
radiometrically dated.  Our objectives were first to complete the
paleomagnetic record with paleointensity determinations and to
identify this geomagnetic event by radiometric dating.  We also wanted
to check the presence of excursional directions.  In this paper we
report directional results and high quality paleointensity
determinations for four lava flows from Amsterdam Island.  Theses
results combined with \chemical{^{40}Ar/^{39}Ar} isotopic ages provide
more concrete evidence for the occurrence of a geomagnetic excursion
recorded in Amsterdam lava during the late Pleistocene which may
corresponds either to Mono Lake or Laschamp.

\section{Geological setting and Paleomagnetic sampling}

\begin{figure}
\includegraphics{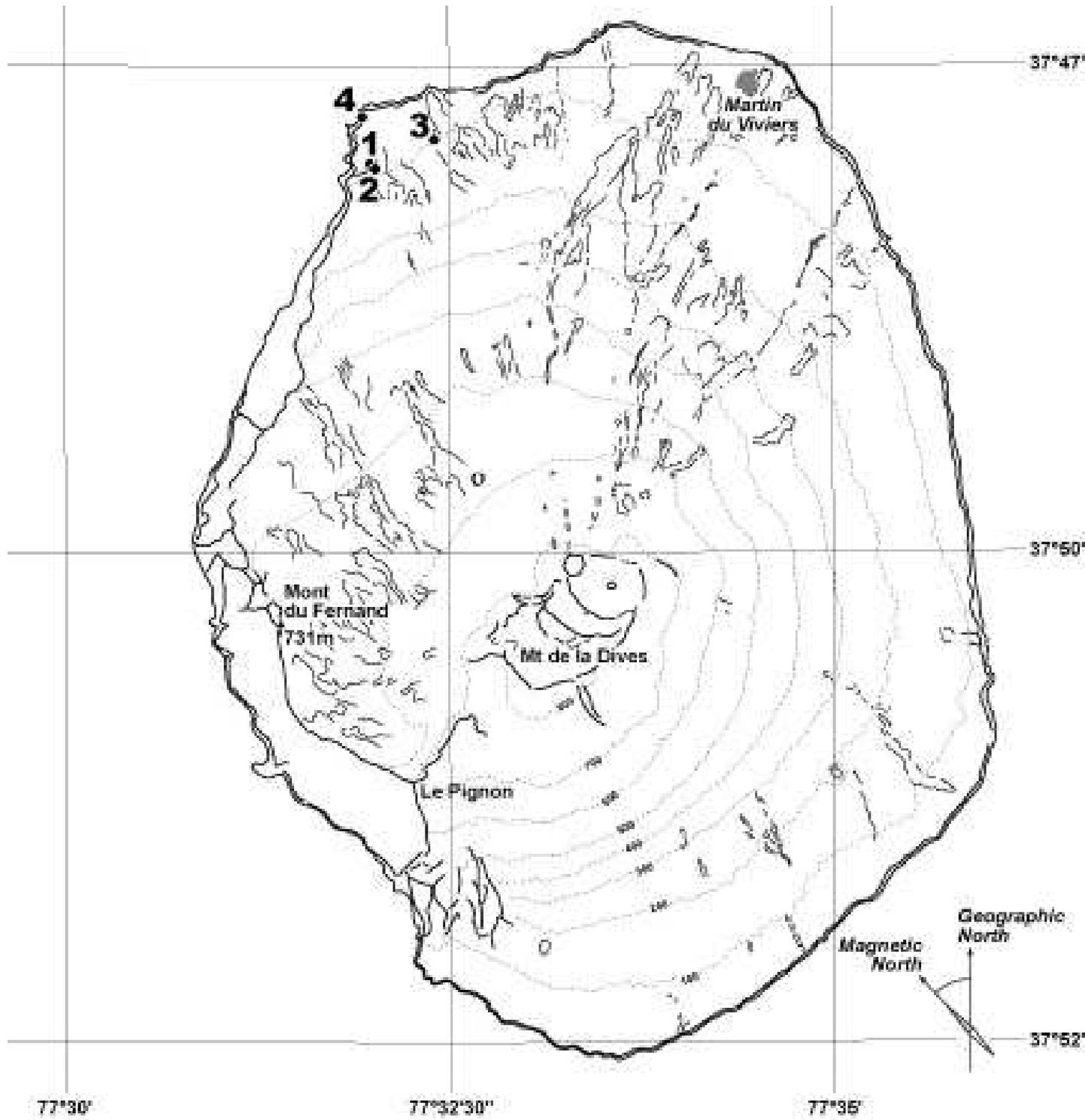}
\caption{Location of the four sampled flows on a topographic map of Amsterdam Island}
\label{map}
\end{figure}

Amsterdam Island stands where a mantle plume is interacting with a
migrating mid-ocean ridge \cite{Small95}.  Amsterdam/Saint-Paul
plateau was built within the past 4 Ma since the South East Indian
Ridge (SEIR) passed over the Amstardam/Saint-Paul hotspot
\cite{Royer88}.  Currently, Amsterdam Island (37.5\degr S, 77.3\degr
E) is located 50 km southwest of the SEIR, which is migrating to the
northeast away from the plume. Lying on the nearly stationary
Antarctic plate, it has been suggested that the Amsterdam/Saint-Paul
plateau is still forming \cite{Graham99} and that volcanic risk still
exists at Amsterdam and Saint Paul Islands \cite{Johnson00}.  Thus
located in the immediate vicinity of a mid-ocean ridge, lava from
Amsterdam has mainly tholeiitic geochemical composition, ranging from
olivine tholeiite to plagioclase basalt.  From a structural point of
view, this volcanic island corresponds to a double strato-volcano
\cite{Gunn71}.  The first cycle of activity is represented by a
volcano centered on the southern part of the island, where Mount du
Fernand and Le Pinion correspond to remnants of a volcanic caldera
(Fig.~\ref{map}).  The west flank of this early volcano collapsed
along two main faults oriented NW-SE and NE-SW, respectively, creating
about 800 m vertical offset and an unreachable escarpment that exposes
a multitude of thin lava flows.  The second cycle of activity built
the Mount de la Dives volcano, which is a simple almost symmetrical
volcanic cone rising to 881 m above sea level. This episode represents
a displacement of the eruptive center of 2 km toward the
east-northeast.  The caldera on top of this young cone preserves
evidence of a lava lake filled by several episodes of lava effusion
which overspilled as voluminous, pahoehoe lava flows.  The flanks of
this volcano have regular slopes, 7 to 13\degr, and are not dissected
by erosion.  Marine erosion has formed a cliff 25-50 m high at sea
level around the island which makes access from the sea difficult.
Although radiometric ages were not available before the present study,
the nearly-pristine morphology leaves no doubt that this volcanic
activity is recent.  Some parasitic cinder cones exist on the lower
flanks of Mount de la Dives volcano.  The youngest one is supposed to
be formed during the last century.

The paleomagnetic samples were taken in few hours during the resupply
of the permanent scientific station by the Marion-Dufresne vessel.
Because of the limited schedule, we decided to confine the sampling to
the northwestern coast where Watkins \& Nougier \shortcite{Watkins73}
studied six lava flows in a restricted area.  Four of these flows
provided excursional directions.  We grouped Watkins \& Nougier's
\shortcite{Watkins73} flows 17 and 18, because of their confusing
boundary geometries, into a single unit (am1).  Probably they
constitute a compound lava flow corresponding to a single volcanic
eruption.  Their flow 19, herein called am2, is located several meters
above and corresponds undoubtly to a different volcanic eruption.
Unfortunately, we were not able to reach Watkins \& Nougier's (1973)
flows 13, 14, and 15, because of the presence in this sector of
numerous fur seals which made the access dangerous.  Instead, we
sampled two previously-unstudied flows (am3 and am4; Fig.~\ref{map}).
Seven cores from each of flows am1-2-4 and eight from flow am3 were
drilled using a gasoline-powered portable drill and were oriented with
respect to geographic north by means of both solar sightings and
magnetic compass plus a clinometer.  Based on stratigraphic
relationships, all four sampled volcanic units belong to the youngest
flows of the Mount de la Dives volcano.  On the basis of field
observations, flow am4 is certainly younger than the other three. We
have, on the other hand, no way to decipher the age relations between
flows am1-2 and flow am3.

\section{Paleodirection determinations}

\subsection{Experimental procedure} 

\begin{figure}
\includegraphics{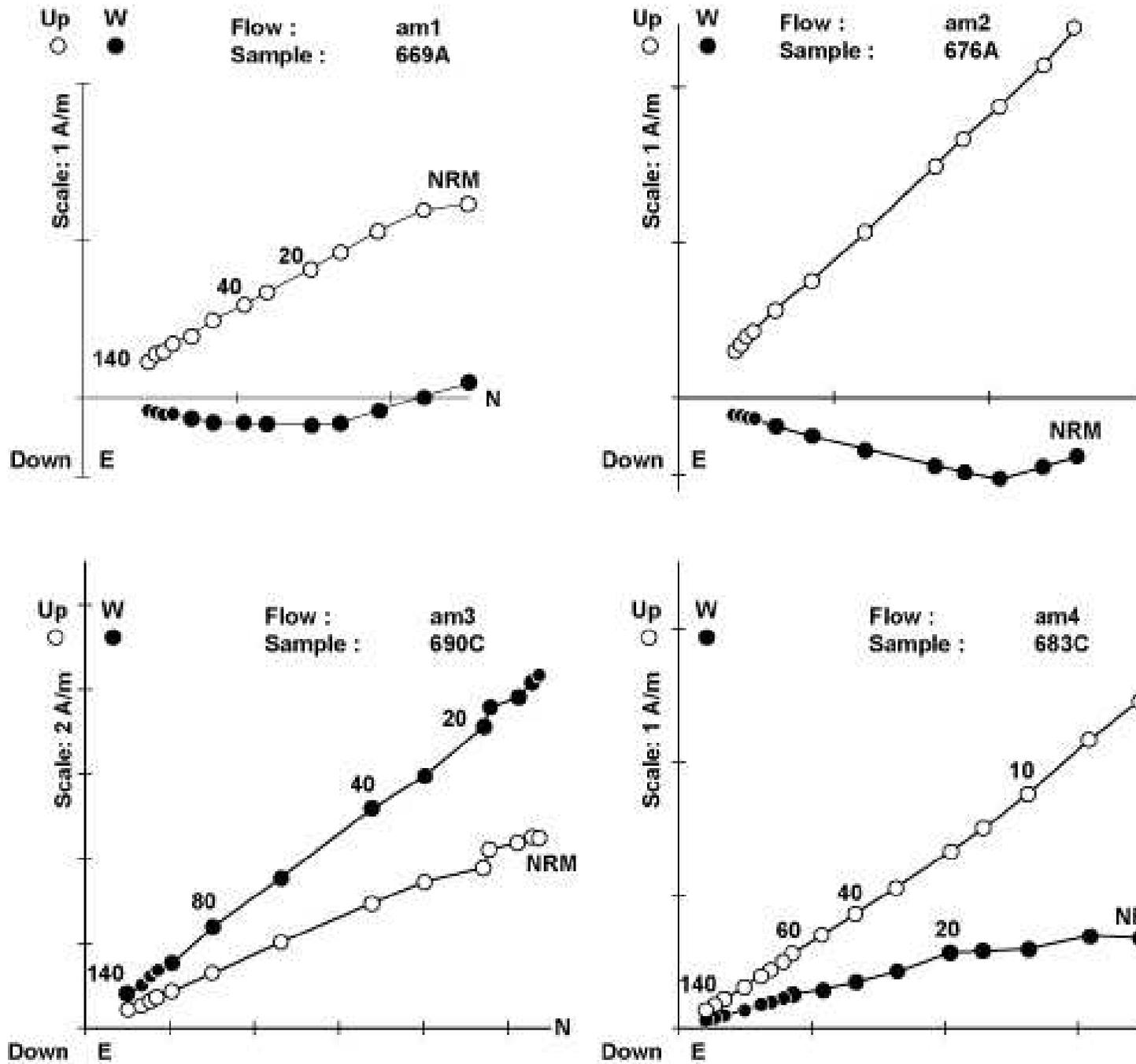}
\caption{Orthogonal projections of alternating-field
  demagnetization for one pilot sample from each lava flow. Solid
  (open) symbols represent projection into horizontal (vertical)
  planes.}
\label{zijd}
\end{figure}

For the analysis of remanence direction, we first treated a pilot
sample from each flow using a detailed experimental procedure
involving up to 13 alternating fields (AF) cleaning steps in order to
check the possible presence of unstable components of remanence.
Because of the simple behavior of remanence upon cleaning
(Fig.~\ref{zijd}), we used only 4 or 5 AF steps for the remaining
samples.  Measurements of remanent magnetization were carried out with
a JR-5A spinner magnetometer and the AF treatments, with a laboratory
built AF demagnetizer in which the sample is stationary and subjected
to peak fields up to 140 mT.  The analysis of the demagnetization
diagrams is straightforward for all the samples but one, sample 693
from flows am3 was contaminated by a significant parasitic
magnetization of unknown origin. We determined the Characteristic
Remanent Magnetization (ChRM) by means of the principal component
analysis using the Maximum Angular Deviation (MAD) \cite{Kirschvink80}
as a measure of the inherent scatter in directions. In order to check
if the principal component is a robust estimate of the sample ChRM, we
compared this direction with the fitting line constrained through the
origin. When the angle between these two directions exceeds the MAD
\cite{Audunsson97} we concluded that the principal component is
statistically different from the ChRM, and thus that the ChRM is not
perfectly isolated. This method led to the rejection of only one
sample (693) from further analysis, considering that no ChRM could be
successfully determined for this sample. We averaged the directions
thus obtained by flow, and calculated the statistical parameters
assuming a fisherian distribution (Table~\ref{directions}). The ChRM
directions are well clustered in each flow with rather small values of
the 95 \% confidence cone about the mean direction ($\alpha_{95}$),
all $\le$ 5\degr.

\subsection{Paleodirection results}

The directional results obtained for flows am1 and am2 corroborate
exactly the finding by Watkins \& Nougier (1973): an intermediate
polarity with associated Virtual Geomagnetic Poles (VGPs) close to the
equator is found. This result is not surprising insofar as the
remanence of the lava from Amsterdam is not contaminated by
significant spurious components. This was not the case, for example,
in a recent study on the volcanic sequence from Possession Island
\cite{Camps01}, where the authors concluded that the intermediate
directions initially described by Watkins et al. (1972) correspond to
reversed directions which had been incompletely cleaned of their
present-day field viscous overprint. Here, we believe that the
previously published data for Amsterdam Island \cite{Watkins73}, which
were not resampled in the present
study are, equally reliable.\\

\begin{figure}
\includegraphics{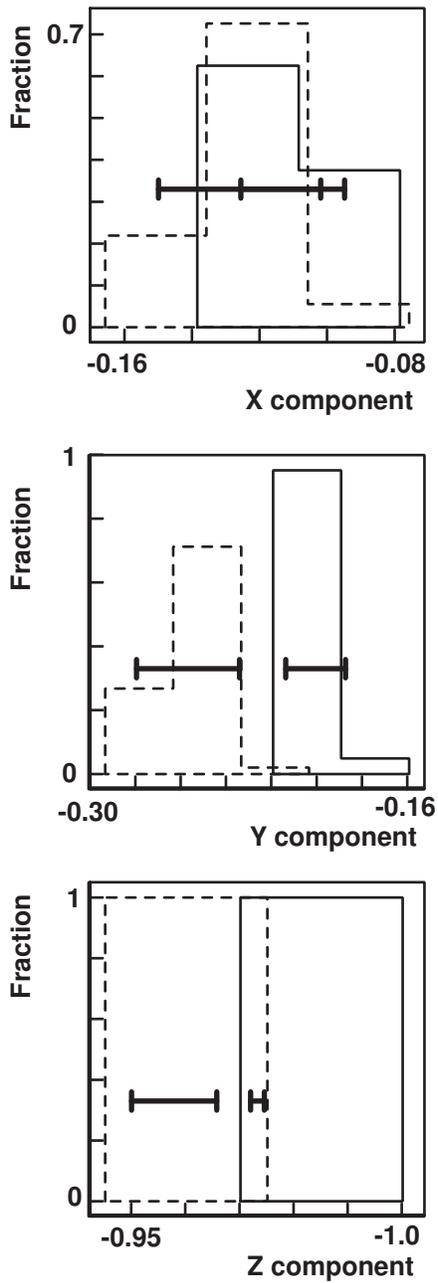}
\caption{The boostrap test for a common mean direction for flows am1 and am2 (Tauxe, 1998) illustrated by 
  the histograms of the cartesian coordinates of the bootstrapped
  means for flows am1 (solid line) and am2 (dashed line). Because the
  95\% confidence intervals for Y and Z components do not overlap, we
  assume that the two lava flows have a significantly different mean
  direction.}
\label{boot-mean}
\end{figure}

Because flows am1 and am2 yield two similar directions and because
they are in a single sequence on a small cliff, one can ask whether
the time elapsed between these two flows is long enough to consider
them separately. To try to reply briefly, we performed the bootstrap
test for a common mean \cite{Tauxe98}. Because the 95 \% bound
interval for the Cartesian Y and Z coordinates calculated for these
two directions do not overlap each other (Fig.~\ref{boot-mean}), we
concluded that these directions are statistically different and thus
can be analyzed individually.  Flows am3 and am4 give normal
directions. They can be used to complete Watkins \&
Nougier's~\shortcite{Watkins73} dataset in order to estimate the
amplitude of secular variation from the Indian Ocean for the Brunhes
period.

\section{Paleointensity determinations}

\subsection{Experimental procedure}

Paleointensity determinations were carried out using the classical
Thellier and Thellier ~\shortcite{Thellier59} method.  The samples are
heated two times at each temperature step, in the presence of a field
positive for the first heating and negative for the second heating.
Partial thermoremanent magnetization (pTRM) checks were performed
every two steps in order to detect magnetic changes during heating.
We used a laboratory field of 30 $\mathrm{\mu T}$ aligned along the
core z axis.  All heatings and coolings were done in a vacuum better
than $\mathrm{10^{-4}}$ mbar with the intention of reducing
mineralogical changes during heatings, which are usually due to
oxidation.  Samples were heated in 14 steps between 150 and 580
$\degr$C, by 50$\degr$C between 150 and 500 $\degr$C, then 20$\degr$C
steps between 500 and 540$\degr$C, and finally 10$\degr$C steps
between 540 and 580$\degr$C. Each heating-cooling step required about
10 hours.  Because paleointensity measurements require time-consuming
procedures, it is important to detect unsuitable samples before
carrying out the full experiments.

\subsection{Sample selection and rock magnetism properties}

\begin{figure}
\includegraphics{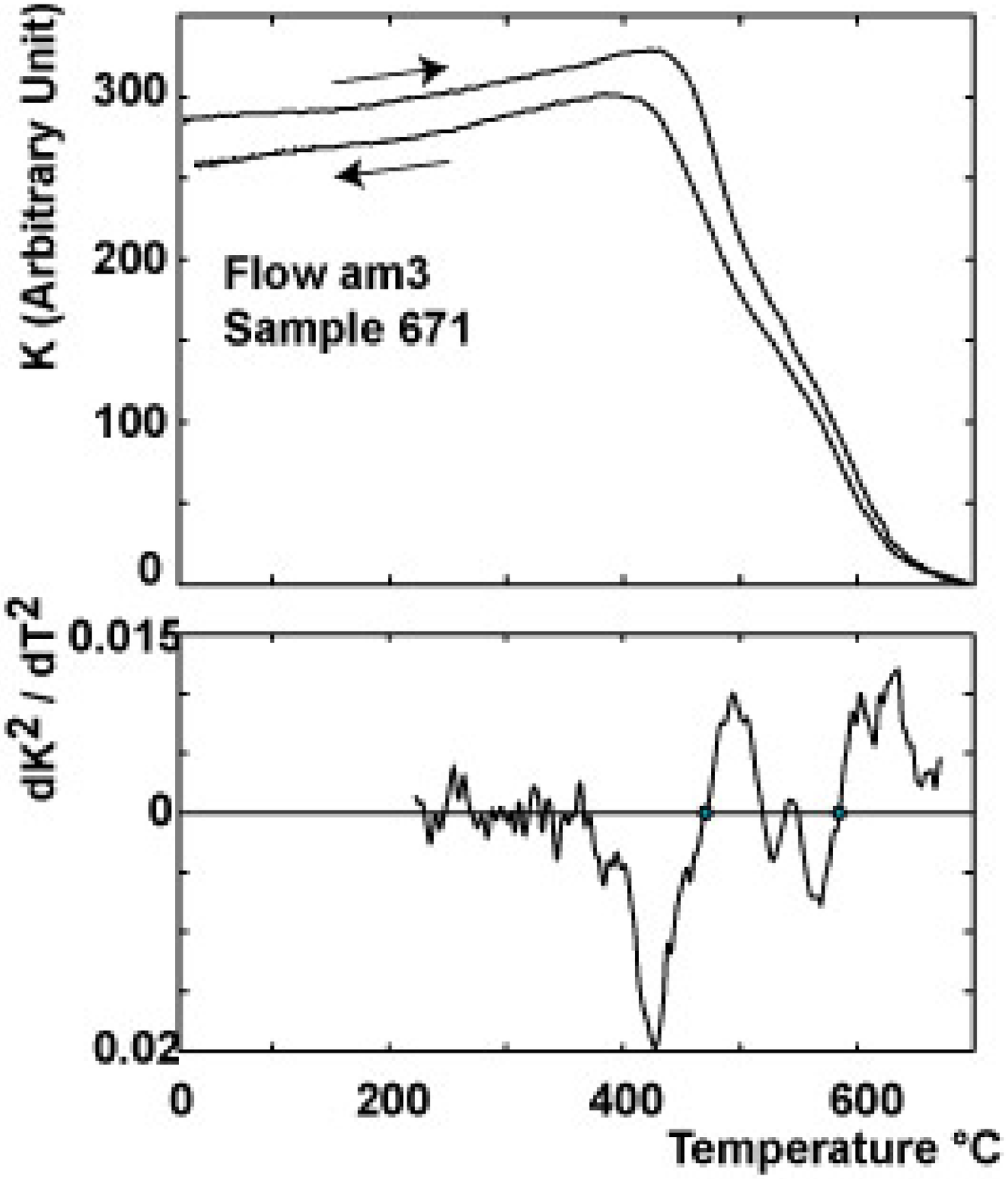}
\caption{Example of thermal variation of weak field magnetic susceptibility K (measured in induction
  B equal to 100$\mu$T) against temperature T(\degr C) showing a good
  reversibility and second derivative of the smoothed data of the
  heating curve.  The mean Curie temperature is defined when the
  second derivative increases to zero (Pr\'evot et al., 1983)}
\label{kt}
\end{figure}

Volcanic rocks used for absolute paleointensity determinations must
satisfy the following conditions :

\begin{enumerate}
  
\item The natural remanent magnetization (NRM) of samples must consist
  of a single component close to the mean characteristic remanence
  direction of the flow.  In addition, the viscosity
  index~\cite{Thellier44} must be small enough to obtain reliable data
  during the heating demagnetization steps at low temperatures.  We
  found that all the samples have magnetic viscosity coefficients
  smaller than 4\%; therefore no samples were eliminated on this
  basis, except sample 693 which shows a strong secondary component of
  magnetization.
  
\item The magnetic properties of the samples must be thermally stable.
  To check it, we performed continuous low-field magnetic
  susceptibility measurements under vacuum (better than 10$^{-2}$
  mbar) as a function of temperature (Fig.~\ref{kt}) for the 28
  remaining samples.  The device used for this experiment is a
  Bartington susceptibility meter MS2 equipped with a furnace in which
  the heating and cooling rates remain constant at 7$\degr/$mn.  Seven
  samples having irreversible thermomagnetic curves were rejected for
  the final selection of paleointensity results. Mean Curie
  temperatures~\cite{Prevot83} vary between 450 and 570\degr C
  (Table~\ref{a-parameter}). These values also indicate that the
  magnetic carriers are mainly Ti-poor titanomagnetite (x $<$
  0.2)~\cite{Dunlop97}.
  
\item The remanence carriers must be single domain (SD) or
  pseudo-single domain (PSD) grains.  It is widely accepted that multi
  domain (MD) grains give erroneous results because of the inequality
  of their blocking and unblocking temperatures and the influence of
  the thermal prehistory on pTRM intensity~\cite{Vinogradov89}.  In
  order to determine the domain structure, we measured the hysteresis
  parameters using an alternative gradient force magnetometer at the
  Universidad Nacional Autonoma de Mexico.  According to the criteria
  defined by Day et al.~\shortcite{Day77} all the hysteresis
  parameters are in the PSD part of the plot (Fig.~\ref{day-plot} and
  Table~\ref{a-parameter}).  However, a mixture of SD and MD grains
  could give the same result.  Most samples are characterized by a
  very high median destructive field (Table~\ref{a-parameter}) which
  is a further evidence of the presence of SD-PSD grains.

\end{enumerate}

\subsection{Preliminary selection of paleointensity data}

The parameters used as criteria for a preliminary data selection are
defined as follows, based on selection criteria commonly used for
paleointensity experiments.

\begin{enumerate}
  
\item The number $N$ of successive points on the linear segments
  chosen to calculate the paleointensity must be at least 4.
  
\item The fraction of NRM destroyed on this segment must be greater
  than 1/3.
  
\item The MAD calculated with the principal component calculated in
  the temperature interval used for paleointensity estimate must be
  less than 15\degr ~and the angle $\alpha$ between the vector average
  and this principal component also less than 15\degr
  ~\cite{Selkin00}.
  
\item The pTRM checks have to be positive, i.e. the deviation of pTRM
  quantified by the difference ratio ~\cite{Selkin00}, which
  corresponds to the maximum difference between repeat pTRM steps
  normalized by the length of the selected NRM-pTRM segment, has to be
  less than 10\% before and within the linear segment.  Failure of a
  pTRM check is an indication of irreversible magnetic and/or chemical
  changes in the ferromagnetic minerals during the laboratory heating.

\end{enumerate}

\subsection{PTRM-tail test}

In order to give us some indication about the domain states as a
function of the temperature, we performed the pTRM-tail test that was
first introduced by Bol'shakov \& Shcherbakova~\shortcite{Bolshakov79}
and modified later by Shcherbakova et al.~\shortcite{Shcherbakova00}.
The principle of this test is as follows: if a sample is given a pTRM
on an interval [T$_1$, T$_2$] ($\mathrm{T_1 > T_2}$), then is heated
up to T$_1$ and cooled down in zero-field, the pTRM will be completely
demagnetized only if the remanence carrier is single-domain. For
pseudo-single-domain or multi-domain material, the pTRM will be
completely demagnetized only after heating to a temperature higher
than T$_1$, this temperature reaching all the way to T$_C$ for MD
grains~\cite{Bolshakov79}.  Note that this test can only be applied to
samples that do not alter chemically during heating.  As discussed
later, the Amsterdam samples are unusually stable, permitting wide
application of this test.

The pTRM-tail test was carried out using a thermal vibrating
magnetometer which allows the measurement of the magnetic remanence
and the induced magnetization of a rock sample. The dimensions of the
sample are 11 mm height and 10 mm diameter. The static residual field
in the heating zone is less than 20 nT. It is possible to apply a
direct field on the sample by sending a constant current to an inner
coil placed between the detection coils and the heater. The two
detection coils are connected in opposition. The sample is
alternatively translated from the center of the first detection coil
to the center of the other detection coil with a frequency of 13.7 Hz
with 25.4 mm amplitude.  The heater is powered with an alternating
pulse width modulated current at a frequency of 3740 Hz. The output
signal is directly applied to the current input of the lock-in
amplifier Stanford Research SR830. After signal acquisition and
calibration, we measure the magnetization moment versus temperature
with a precision of $2 \times 10^{-8} \mathrm{Am^{2}}$ (with a time
constant of 300 mS).

\begin{figure}
\includegraphics{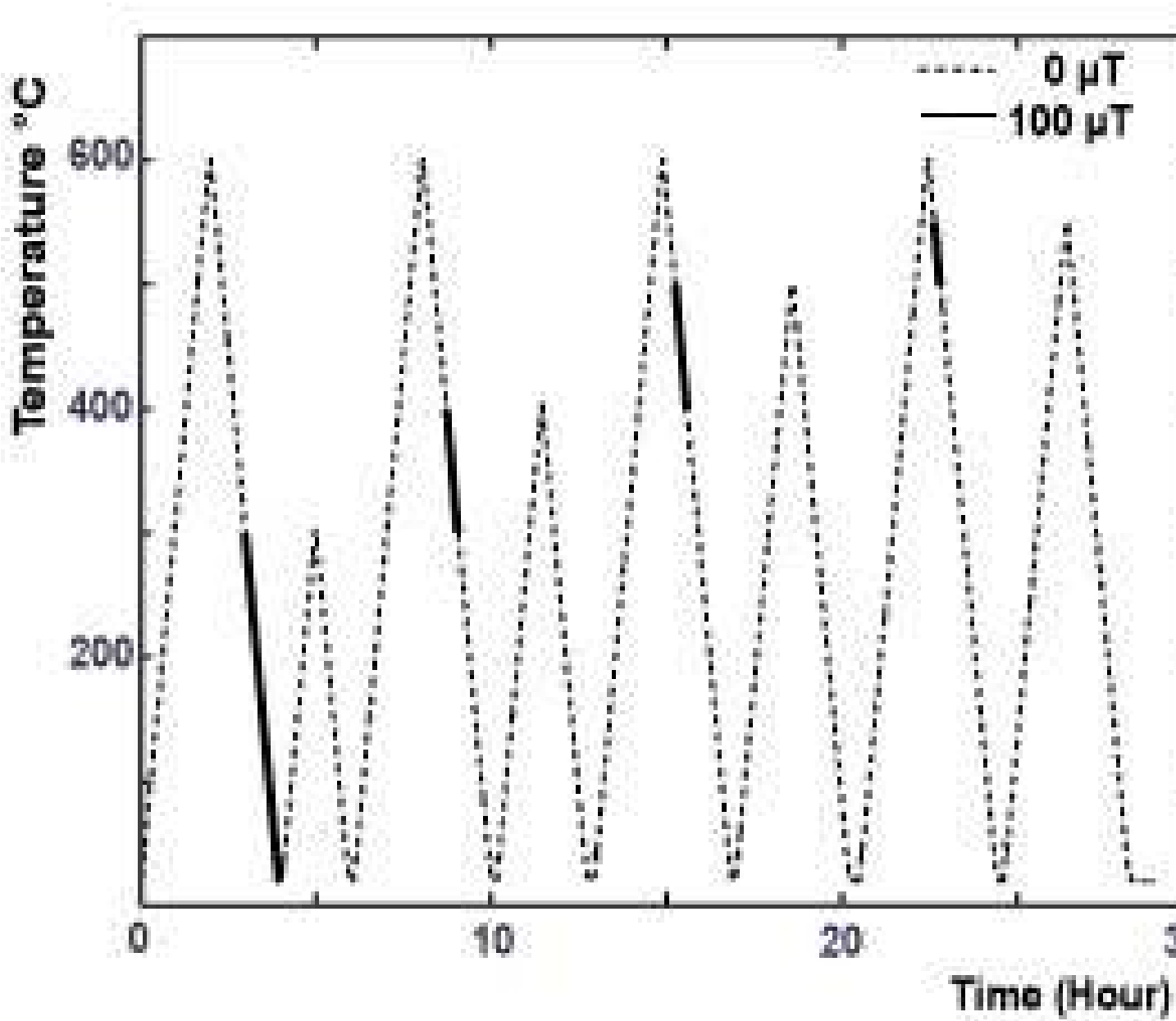}
\caption{Temperature as a function of time for a sample heated and cooled at a constant rate of 5\degr /mn.
  The curve shows the succession of pTRM acquisitions in an applied
  field of 100 $\mu$T (solid line) during cooling and demagnetizations
  (dashed line) carried out during the pTRM-tail test. Parameter A is
  calculated as the ratio of the intensity of the tail of pTRM
  normalized by the original pTRM, both measured at room temperature}
\label{tail-test}
\end{figure}

\begin{figure}
\includegraphics{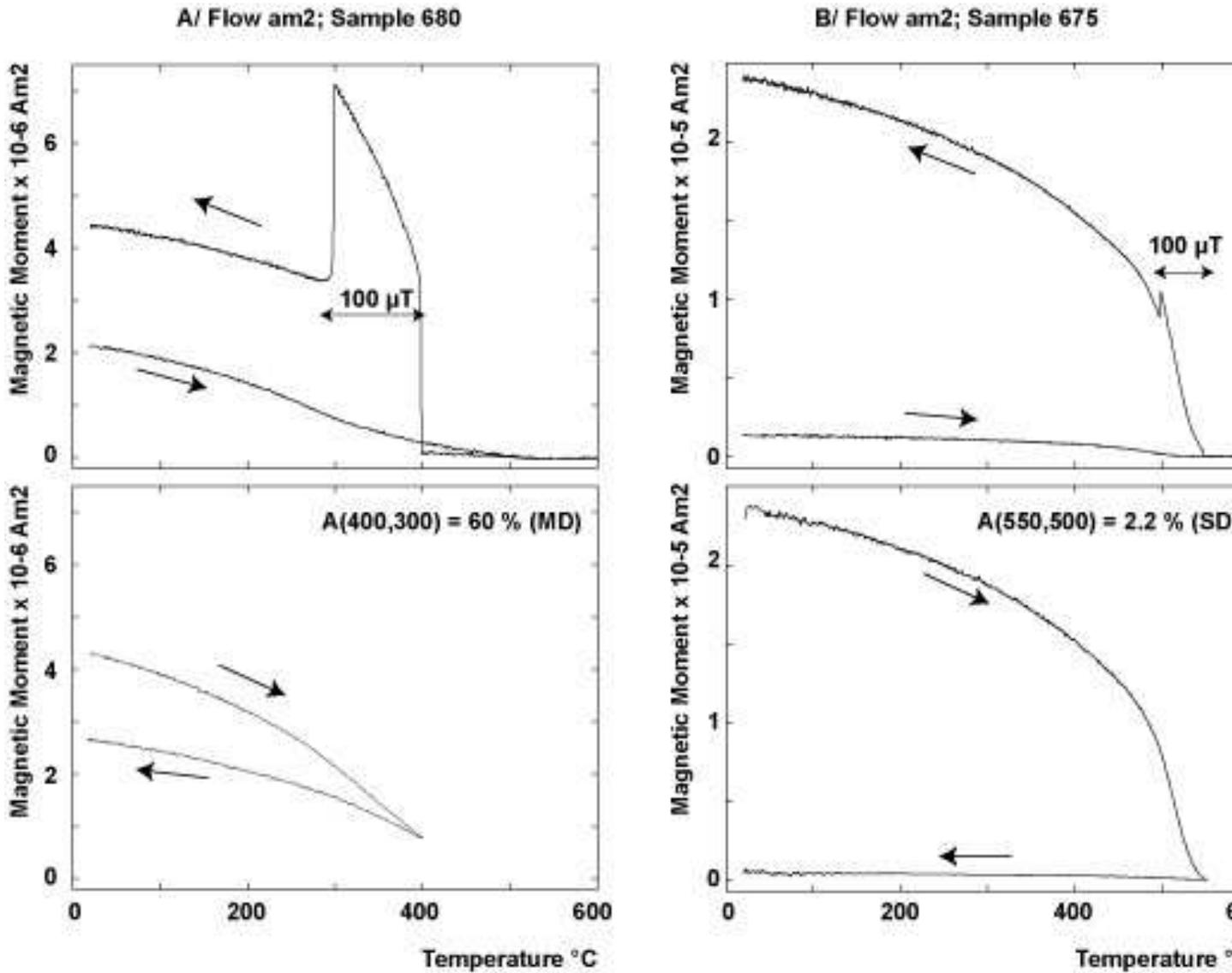}
\caption{Thermomagnetic curves acquired during pTRM acquisition (top) and pTRM
  demagnetization (bottom). In example A the pTRM is imparted on the
  interval (400\degr C-300\degr C). Subsequent demagnetization to
  400\degr C leaves an important tail. In example B, the pTRM is
  imparted on the interval (550\degr -500\degr C) and is almost
  completely demagnetized after heating back to 550\degr C.}
\label{tail-resultat}
\end{figure}

PTRM acquisitions were performed in air on sister samples (i.e.,
adjoining samples from the same paleomagnetic core), using a 100
$\mu$T field, in four different temperature intervals: [300\degr C,
T$_{room}$], [400\degr C, 300\degr C], [500\degr C, 400\degr C] and
[550\degr C, 500\degr C]. PTRMs were imparted "from above": the
samples were first demagnetized by heating them in zero field to
T$_C$, then a 100 $\mu$T is applied during the cooling down between
the temperatures T$_1$ and T$_2$, and pTRM(T$_1$,T$_2$) was measured
at room temperature.  Samples were subsequently heated again to T$_1$,
cooled down in zero field, and the tail of pTRM(T$_1$,T$_2$) measured
at room temperature.  Fig.~\ref{tail-test} illustrates the succession
of pTRM acquisitions and demagnetizations.  We calculated the
parameter A defined by
\begin{equation}
\mathrm{A(T_1,T_2) = \frac{tail[pTRM(T_1,T_2)]}{pTRM(T_1,T_2)} 100 \%}
\end{equation} 
as the relative intensity measured at room temperature of the pTRM
tail remaining after heating to T$_1$.  According to the criteria
defined by Shcherbakova et al. (2000), $\mathrm{A(T_1, T_2) < 4\%}$
corresponds to SD, $4\% < \mathrm{A(T_1, T_2)} < 15\%$ to PSD and
$\mathrm{A(T_1, T_2)}$$>$15\% to MD.  Table~\ref{a-parameter} shows
the values of $\mathrm{A(T_1, T_2)}$ for the four temperature
intervals used on the 17 remaining samples.  All the samples have an
MD response for pTRM's imparted in the intervals [300\degr C, T$_r$]
and [400\degr C, 300\degr C].  However they all have a PSD response
(except sample 669 which has an MD response) for the pTRM given in the
interval [500\degr C, 400\degr C], and PSD or SD response for a pTRM
given in the interval [550\degr C, 500\degr C].
Fig.~\ref{tail-resultat} illustrates typical MD and SD thermomagnetic
behavior.  Guided by these results we did not include any point on the
Arai plot acquired before 400\degr C (500\degr C for sample 669) to
calculate the paleointensity estimates.

\subsection{Paleointensity results}

\begin{figure}
\includegraphics{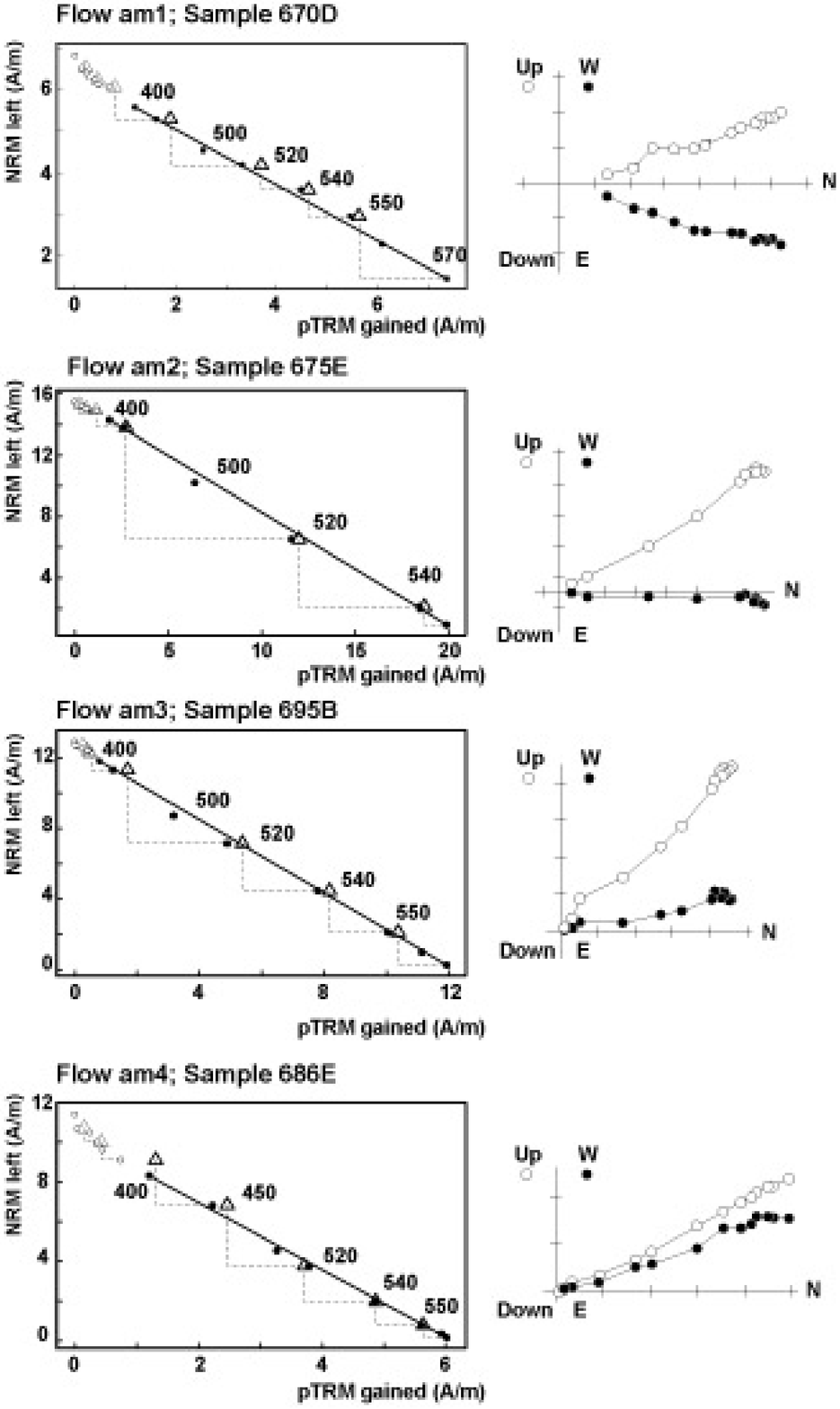}
\caption{Examples of typical Arai plots with the corresponding orthogonal
  vector projections. On the Arai plots the triangles represent the
  pTRM checks and solid (open) symbols correspond to accepted
  (rejected) points. In the orthogonal vector diagrams, solid (open)
  cercles represent projection into the horizontal (vertical) plane.}
\label{arai}
\end{figure}

Results are plotted as NRM lost as a function of pTRM gained on an
Arai graph~\cite{Nagata63}.  Examples of typical good samples are
shown on Fig.~\ref{arai} with associated orthogonal vector diagrams.
Fourteen samples fulfilled all the criteria defined above and were
then considered as yielding reliable results (Table~\ref{table-pi}).
Most results have high quality factor ($q$) values (between 15 and 70)
and the success rate of 50\%, calculated on the whole collection, is
very high and somewhat unusual for natural rocks.

Averaging four acceptable results, the flow am1 (excursional unit)
gives a paleointensity of 24.6 $\mu$T and a Virtual Dipole Moment
(VDM) of 3.7$\times$10$^{22}$ $\mathrm{Am^2}$ (Table~\ref{table-pi}).
The flow am2 (excursional unit) gives an average paleointensity (using
two results) of 24.0 $\mu$T, corresponding to a VDM of
3.4$\times$10$^{22}$ $\mathrm{Am^2}$.  The two normal flows (am3 and
am4) give averages of 32.8$\mu$T (using 5 values) and 46.9$\mu$T
(using 3 values), yielding VDM's of 6.2$\times$10$^{22}$
$\mathrm{Am^2}$ and 7.7$\times$10$^{22}$ $\mathrm{Am^2}$,
respectively.

\section{$^{40}$A\lowercase{r} / $^{39}$A\lowercase{r} age determinations}

\subsection{Analytical procedure}
For each sample, 300 mg of whole rock fragments were carefully
hand-picked under a binocular microscope from crushed 0.5 mm thick
rock slabs. The samples were wrapped in Cu foil to form small packets
as small as possible (11x11 mm.). These packets were stacked up to
form a pile within which packets of flux monitors were inserted every
5 to 10 samples, according to the size of the samples. The stack, put
in an irradiation can, was irradiated, with a Cd shield, for 1 hr at
the McMaster University reactor (Hamilton, Canada) with a total flux
of 1.3 x 1017 n.cm$^{-2}$. The irradiation standard was the
Fish-Canyon sanidine (28.02 Ma; Renne et al.~\shortcite{Renne98}).

The sample arrangement allows monitoring of the flux gradient with a
precision as low as $\pm$0.2 \%.  The step-heating experiment
procedure was described in details by Ruffet et
al.~\shortcite{Ruffet91}.  The mass spectrometer consists of a 120°
M.A.S.S.E.$^{\textcircled{R}}$ tube, a B\"aur
Signer$^{\textcircled{R}}$ source and an SEV 217$^{\textcircled{R}}$
electron-multiplier (total gain: 5$\times 10^{12}$ ) whereas the all
metal extraction and purification lines include two SAES GP50W getters
with St101$^{\textcircled{R}}$ zirconium-aluminium alloy operating at
400\degr C and a -95\degr C cold trap. Samples were incrementally
heated in a molybdenum crucible using a double vacuum high frequency
furnace. The extraction segment of the line was pumped 3 minutes
between each step.

Isotopic measurements are corrected for K and Ca isotopic
interferences and mass discrimination.  All errors are quoted at the
1$\sigma$ level and do not include the errors on the $\mathrm{
  ^{40}Ar^* / ^{39}Ar_K}$ ratio and age of the monitor.  The error in
the \chemical{^{40}Ar^*/^{39}Ar_K} ratio of monitor is included in the
isochron age error bars calculation.

\subsection{$\mathrm{^{40}Ar^*/^{39}Ar_K}$ results}

The very low K- and rather high Ca-contents of the analyzed samples
and their very young apparent age were unfavorable parameters to
produce high quality analyses and to obtain unambiguous results. The
classical calculation method using an \chemical{^{40}Ar/^{36}Ar}
atmospheric ratio measured on an air aliquot resulted in zero apparent
ages for successive degassing steps, probably as a result of an
inadequate procedure for determining this ratio. The very high ratio
of the measured atmospheric to the radiogenic \chemical{^{40}Ar}
favors use of isochron calculation (correlation method:
\chemical{^{36}Ar/^{40}Ar} versus \chemical{^{39}Ar^*/^{40}Ar_K}; e.g.
Turner~\shortcite{Turner71}; Roddick et al.~\shortcite{Roddick80};
Hanes et al.~\shortcite{Hanes85}).  This method does not require "a
priori" knowledge of the measured \chemical{^{40}Ar/^{36}Ar}
atmospheric ratio. Two argon components can be identified using this
calculation method: the first one, related to the atmosphere, is
usually weakly linked to the mineral; the other one, supplied by
radioactive decay of \chemical{^{40}Ar}, is trapped in minerals
structures. The aim of degassing by steps is to separate, at least
partly, these two components, which allows a mixing line to be
defined, the isochron.  In whole rock analysis, a meaningful isochron
must be calculated on a degassing segment which corresponds to the
degassing of a specific mineral phase.

\begin{figure}
\includegraphics{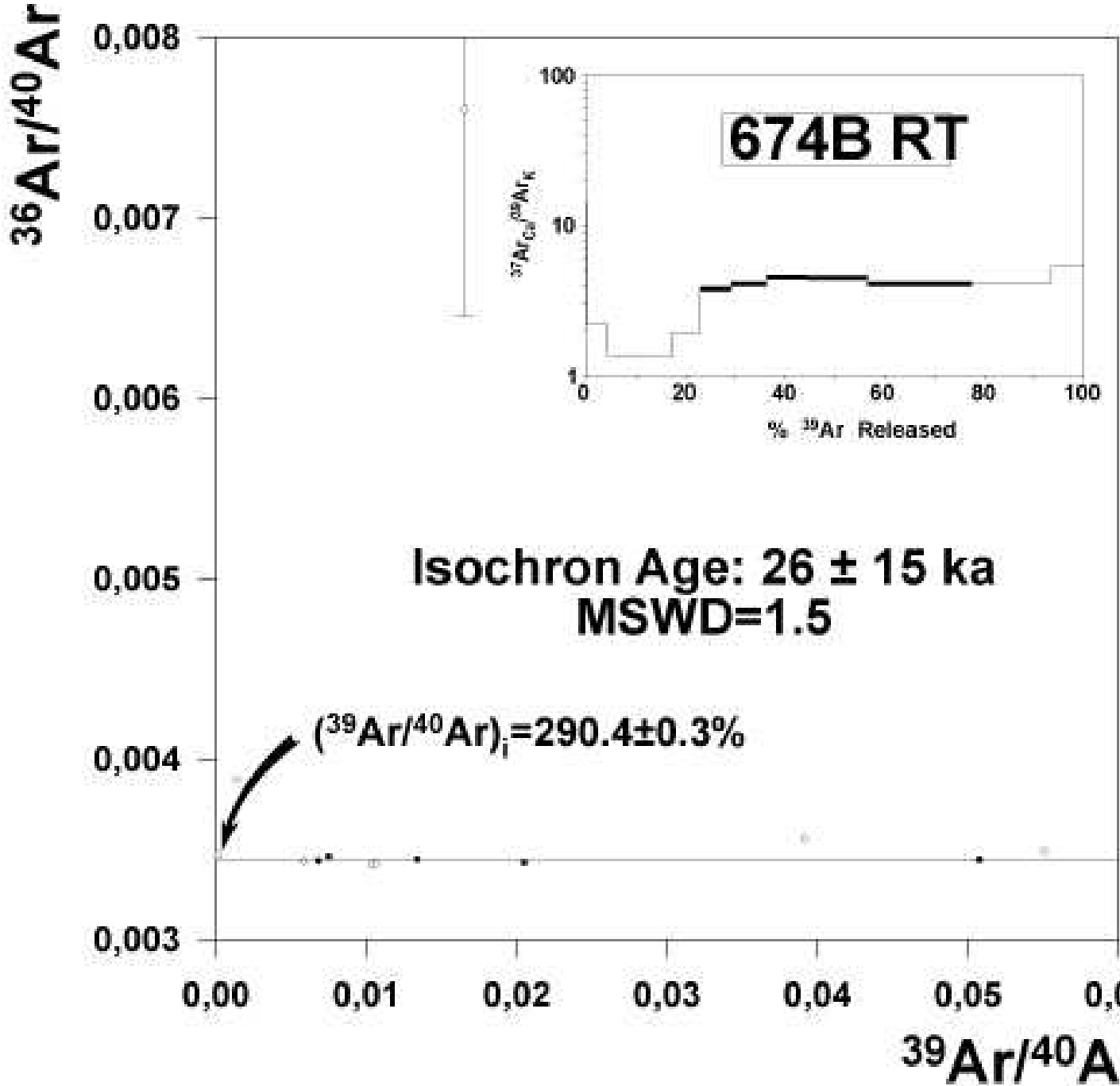}
\caption{Isochron $\mathrm{^{36}Ar / ^{40}Ar}$ versus $\mathrm{^{39}Ar / ^{40}Ar}$ and
  $\mathrm{^{37}Ar_{Ca} / ^{39}Ar_K}$ diagrams of whole rock sample
  674B (Flow am1). The bold line in the Ca/K diagram defines the
  degassing domain used for the isochron calculation.}
\label{age}
\end{figure}

All analyzed samples display rather constant CaO/K$_2$O calculated
ratios in the intermediate temperature range, around 11
(\chemical{CaO/K_{2}O} $=$
\chemical{^{37}Ar_{Ca}/^{39}Ar_K}$\times$2.179; Deckart et
al.~\shortcite{Deckart97}), which suggest degassing of a homogeneous
phase, probably plagioclase as observed in thin sections.  Very young
calculated isochron ages (Fig.~\ref{age}), concordant at the 2 $\sigma$ level, are
obtained from the corresponding degassing steps (Table~\ref{dating}).
They suggest that the two sampled lava flows with intermediate
directions could be as young as 20-25 ka and the lava flow (am3) with
the normal direction slightly older at ca 45 ka.

\section{Discussion}

\subsection{Reliability of paleointensity estimates}
The determination of absolute paleointensity by the Thellier method
imposes many constraints on rock magnetic properties which are often
not respected in natural rocks.  The failure rate is usually around
70-90\%~\cite{Perrin98}.  We obtained results of very good technical
quality for 50\% of the samples in this study.  Nevertheless a recent
study carried out on historical lava flows from Mount Etna showed that
samples which fulfilled all the reliability criteria imposed by the
authors could yield a paleointensity exceeding the real field
paleomagnitude by as much as 25\%~\cite{Calvo02}.  It should never be
forgotten that measurements of paleointensity are only estimates.
Therefore we wish to discuss further the reliability of the
paleointensity measurements performed in the present study.

\begin{figure}
\includegraphics{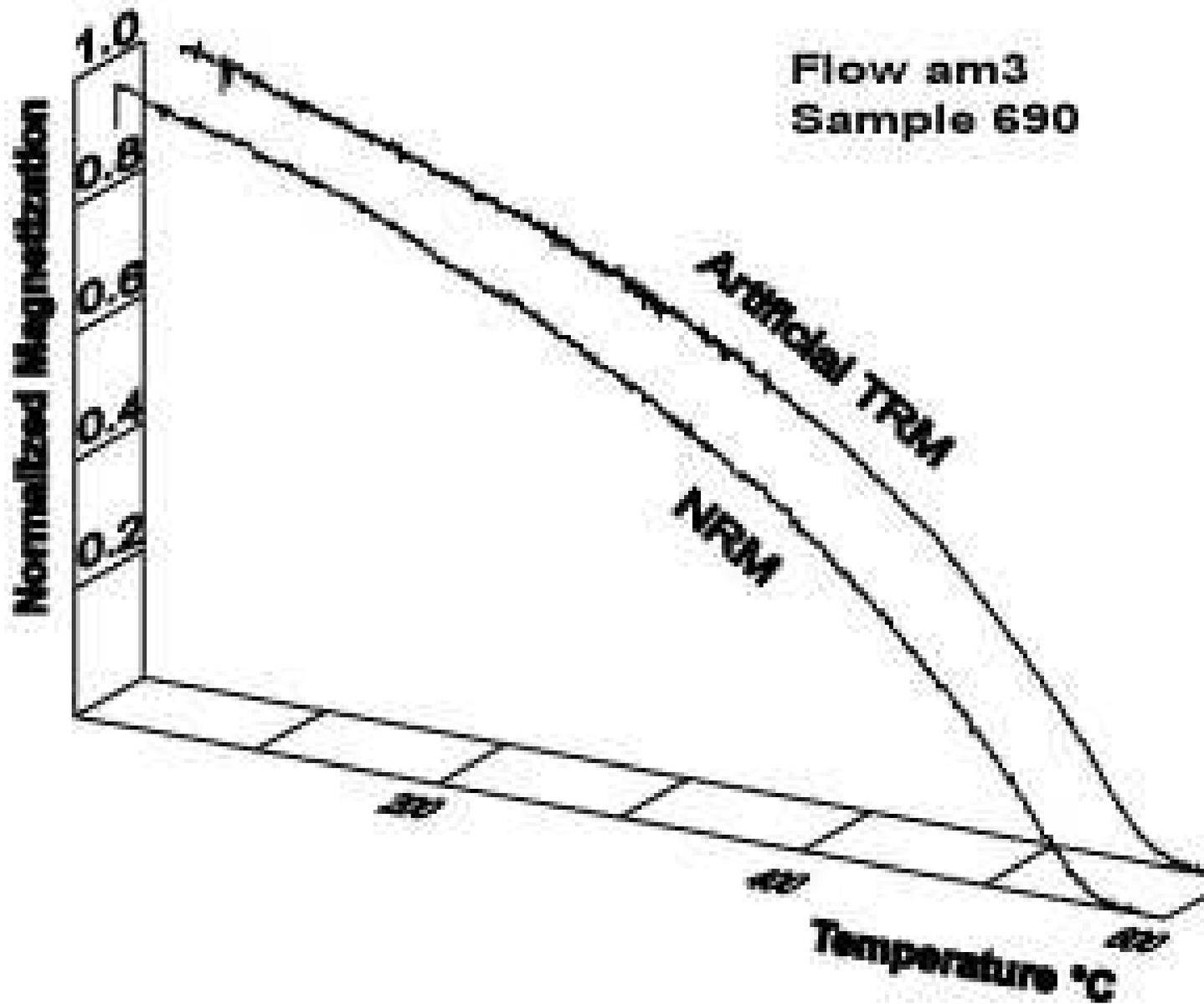}
\caption{Comparison between thermal demagnetization of NRM and artificial total TRM.}
\label{nrm-trm}
\end{figure}

First, we can make sure that the part of NRM used must be a TRM. We
compared continuous thermal demagnetization curves of NRM and
artificial (total) TRM of sister samples measured using the thermal
vibrating magnetometer.  The artificial TRM was imparted in the
direction of the NRM.  Samples have to be drilled in the direction of
the NRM; therefore we did this test only for 3 samples because we did
not have enough material. A result is shown in Fig.~\ref{nrm-trm}.
The remarkable similarity between the thermal demagnetization of the
NRM and of the artificial TRM suggests that the NRM is a TRM and
confirms the thermal stability of Amsterdam lava.

Thermal stability can be further tested independently by comparing the
laboratory Koenigsberger ratios before and after heating.  The ratio
before heating is calculated according to the formula:
\begin{equation}
Q_L = \frac{M_{nrm}}{k_a.H_a},
\end{equation}
where $M_{nrm}$ is the natural remanent magnetization, $k_a$ is the
bulk magnetic susceptibility measured before paleointensity
experiments and $H_a$ is the ancient field obtained from the
paleointensity experiments.  The ratio after heating is defined by:
\begin{equation}
Q'_L = \frac{M_{trm}}{k_b.H_{lab}},
\end{equation}
where $M_{trm}$ is the total TRM obtained by extrapolation of the data
corresponding to the highest temperatures on the Arai plot, $k_b$ is
the bulk magnetic susceptibility obtained after heating and $H_{lab}$
is the laboratory field.  We observed that the two Koenigsberger
ratios have similar values for the samples that gave reliable
paleointensity results (Table~\ref{koenig}). This suggests that these
samples are thermally stable.

\begin{figure}
\includegraphics{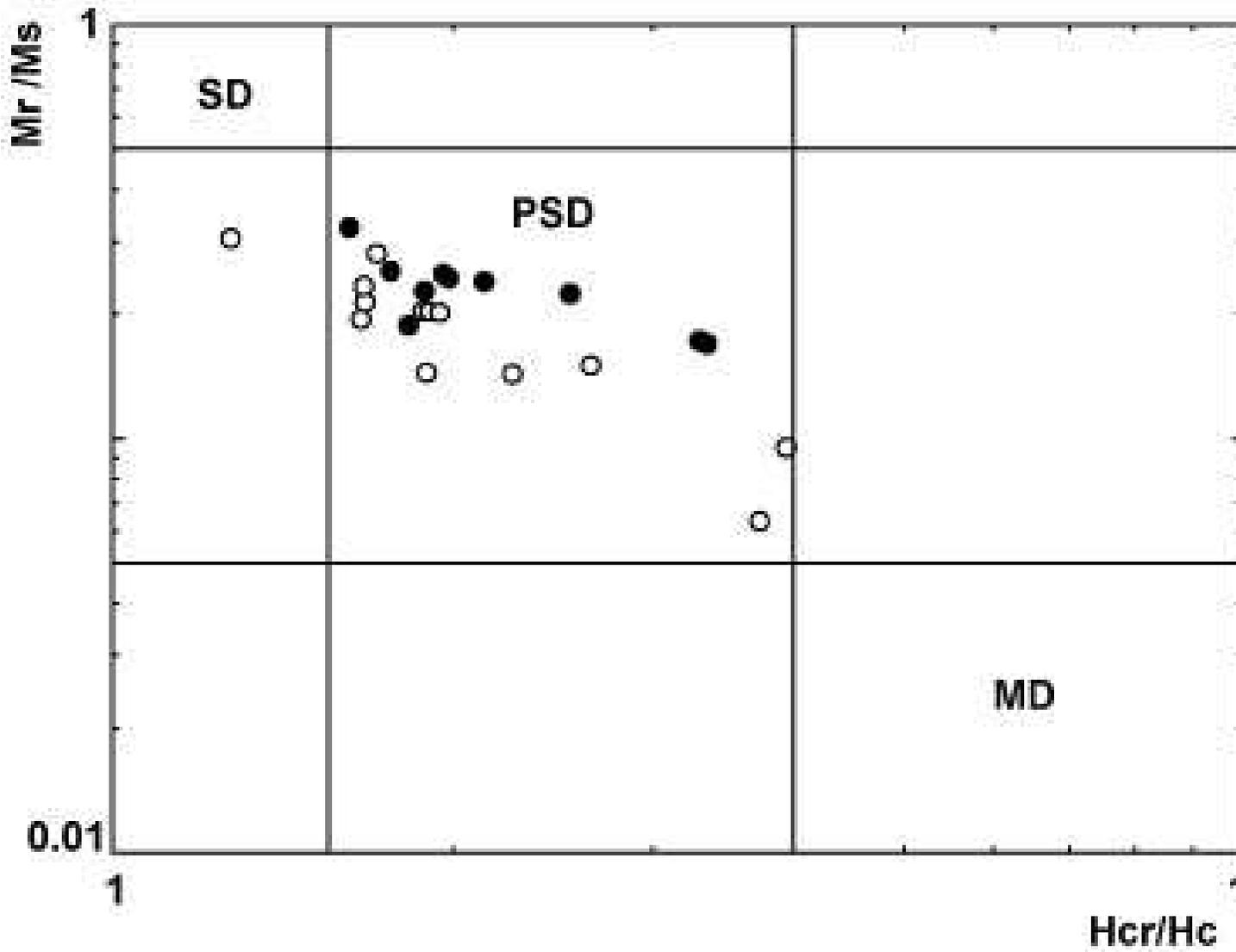}
\caption{Hysteresis parameters ratio measured at room temperature plotted on a log-log scale. Solid (open)
  symbols correspond to accepted (rejected) samples for paleointensity
  experiments on the basis of selection and reliability criteria
  discussed in the text.}
\label{day-plot}
\end{figure}

The hysteresis parameters (Fig.~\ref{day-plot} and
Table~\ref{a-parameter}) show that all the samples have a behavior
characteristic of PSD grains. In natural samples, this is usually
interpreted as the indication of a mixture of SD-PSD and MD grains. We
found that one of the advantages of performing the pTRM-tail-test over
measuring hysteresis parameters is to allow us to discriminate between
MD, PSD and SD thermomagnetic behavior.  This test qualifies directly
the TRM behavior, contrary to the hysteresis curves which make tests
on remanent or induced magnetizations of isothermal origin.  Moreover,
it seems that measurement of hysteresis parameters as a function of
temperature sometimes does not allow detection of changes in domain
structure with heating ~\cite{Carvallo01}.  We note in
Fig.~\ref{day-plot} that the accepted samples (black dots) generally
have higher Mrs/Ms than the rejected samples (white dots).  Thus, on
average the better samples for paleointensity are a little smaller in
effective magnetic grain size.

\subsection{pTRM-tail-test : A new tool for paleointensity experiments ?}
The main methodological originality of the present study is the use of
pTRM-tail tests to select the most suitable portion of the NRM-TRM
diagrams.  Many authors (e.g., Shcherbakova et
al.~\shortcite{Shcherbakova00}) have suggested that recognizing the
multidomain component in the paleointensity experiment is critical for
the exactitude of the final result. The presence of multidomain grains
will invalidate the Thellier method if the unblocking temperature does
not equal the blocking temperature any more. Using the part of the
Arai plot derived from multidomain behavior can thus lead to large
errors in the paleointensity measurement. For example, Shcherbakov \&
Shcherbakova ~\shortcite{Shcherbakov01} showed that, if one were to
ignore the continuous curvature and fit a line to low temperature data
points of synthetic, purely MD samples, one could overestimate the
true value by as much as 60\%.  For the majority of the samples that
we tested, the relative tails A(T$_1$, T$_2$) are very large when pTRM
is imparted at low-temperatures (300\degr C-T$_{room}$ and 400\degr
C-300\degr C), but become smaller and smaller for higher temperature
intervals Table~\ref{a-parameter}, which led us to reject all points
acquired below 400\degr C in calculating the paleointensities.
Shcherbakova et al.~\shortcite{Shcherbakova00} as well as Shcherbakov
\& Shcherbakova~\shortcite{Shcherbakov01} also observed a diminution
of the pTRM tail with increasing temperature intervals, using both
natural and synthetic samples.

\begin{figure}
\includegraphics{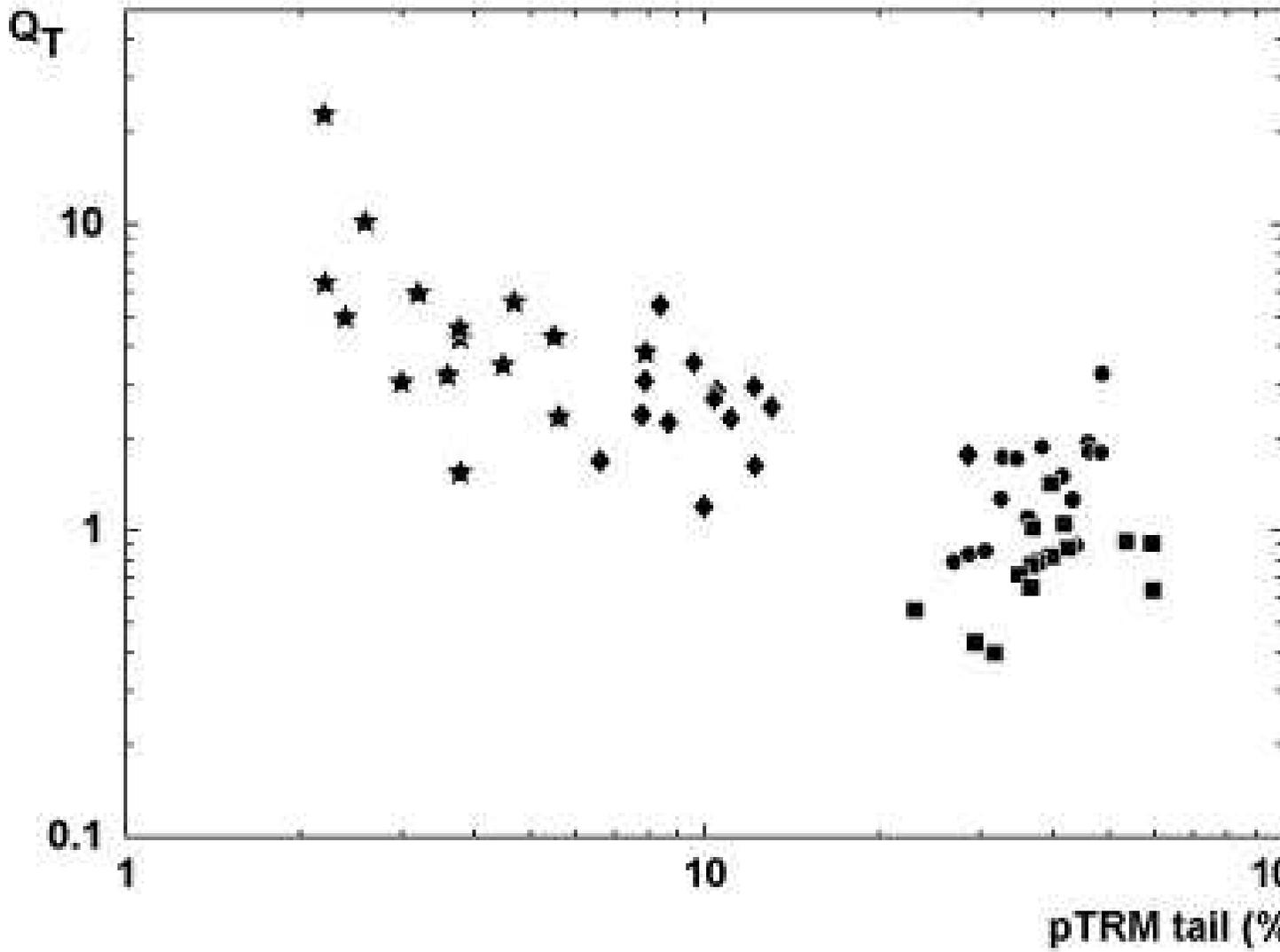}
\caption{Log-Log plot showing the Koenigsberger ratios measured at temperature during the pTRM 
  acquisition when the field is switched off as a function of pTRM
  tail [300,Troom] circles, [400,300] squares, [500,400] diamonds and
  [550,500] stars.}
\label{qt}
\end{figure}

In our case, this behavior could be explained in several ways.
\begin{enumerate}
  
\item Our natural samples are composed of a mixture of grains having
  different sizes. It is possible that the MD part of the remanence
  carriers have lower Curie temperature, whereas PSD and SD grains
  carry the high-temperature remanence~\cite{Dunlop01}.  A simplified
  explanation of what is observed with the variations of A parameters
  is that mainly MD grains are magnetized when they are given a low
  temperature pTRM ($<$ 400\degr C), yielding high A values. For high
  temperature pTRMs ($>$ 400\degr C), mainly PSD/SD grains are
  magnetized and give very small tails.  Another indication of the
  presence of MD material can be extracted from the pTRM acquisition
  curve: when the field is switched off during the cooling, the
  magnetic moment drops because of the presence of induced
  magnetization which is more important for MD than SD grains.
  For example, the pTRM (500\degr C,400\degr C) acquired by sample 680
  has a large tail (60\%), and the magnetic moment drops of about 50\%
  when the field is switched off at 400\degr C
  (Fig.~\ref{tail-resultat}).  But for sample 675, which has a tail
  for the pTRM(550\degr C, 500\degr C) of only 2.2\%, the drop when
  the field is switched off is also much smaller (about 10\% of the
  remanence acquired at this temperature).  This is a general trend
  observed for the ensemble of the pTRM acquisitions-demagnetizations
  (Fig.~\ref{qt}),
although a more precise correlation is difficult to establish.

\item Alternatively, one could argue that for the samples having high Curie points, the high
values of the low temperature pTRM tails might be only due to the fact that the pTRMs acquired
in these intervals are actually very low. The difference between the magnetic moment measured after
acquisition of a small pTRM and the magnetic moment measured after its demagnetization might
then not be significant of any physical process but only reflect an artifact created by the
accuracy of the measurement. 
However, the ranges of magnetization measured are in most cases well above
the sensitivity of the vibrating magnetometer, so we can be quite confident that the measured
values of A in Table~\ref{a-parameter} are physically meaningful. 
\end{enumerate}

From a practical point of view, the pTRM-tail test was not critical for these Amsterdam basalt samples.
Before knowing the A(T$_1$,T$_2$) values, we selected samples and temperature steps for estimating 
paleointensity from our Thellier data using conventional criteria --{\it i.e.}, essentially items i-iv
described above in section 4.3.
The flow-mean values thus obtained 
do not differ significantly from those obtained by adding further data selection according to the 
pTRM-tail test.
In retrospect, this is not unexpected because, the selected samples have 57 to 91\% (on average 81\%)
of their TRM remaining after thermal demagnetization to 400\degr C.
Thus, the pTRM tails of the low-temperature points are too small to have an appreciable effect on the
slope of the best-fit lines of the Arai diagrams, that is, on the paleointensity estimate.
More difficult to understand is the low-temperature slope of sample 680 which have 
more than 30\% of its remanence below 400\degr C.
Its multidomain-size pTRM tails would lead one to expect slope corresponding to large overestimated
of paleointensity, but actually that for 680 is too low by 33\%.
At this time we have no explanation for the departure of this sample from the behavior expected
for multidomain grains \cite{Dunlop01,Shcherbakov01}.
 
\subsection{Implication for the excursional field characteristics}

\begin{figure}
\includegraphics{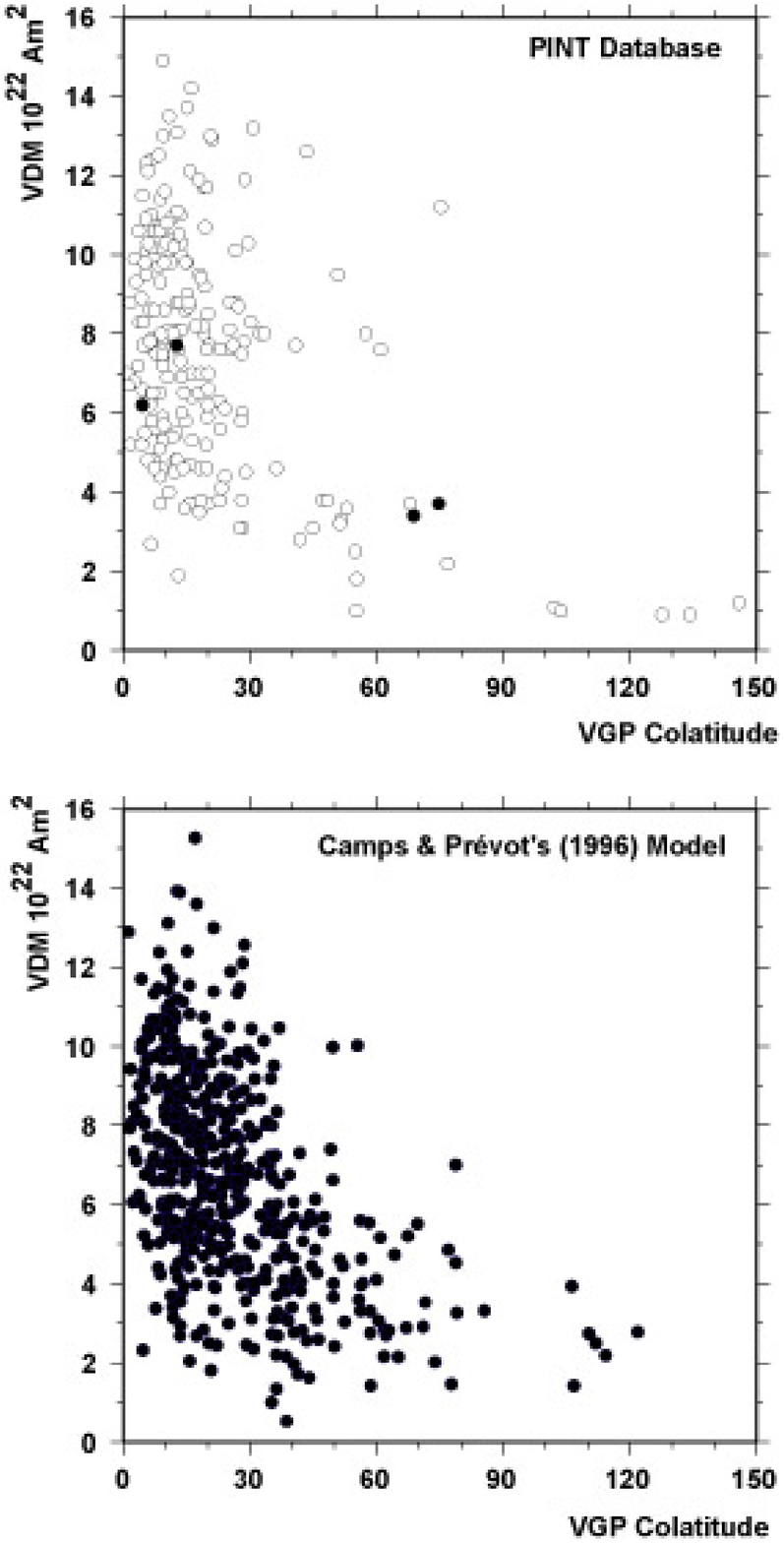}
\caption{a/ VDM for the Brunhes period from the PINT database (open circles) represented as the function of the VGP colatitude and compared to Amsterdam VDMs (Black circles). b/ 500 random simulated VDMs from Camps and Pr\'evot's (1996) statistical model.}
\label{vdm}
\end{figure}

Studies carried out on absolute paleointensity during the Brunhes
period show generally a decrease of the VDM value when the colatitude
of the VGP increases.  This trend is illustrated in Fig.~\ref{vdm}a in
which we gathered Brunhes VDMs from the paleointensity database
PINT2000 using as the unique selection criteria the Thellier
paleointensity method.  The VDM obtained in the present study, showing
approximately a ratio of 2 between the normal and the excursional
field, fit well in this general tendency.  In Fig.~\ref{vdm}b, we have
randomly generated 500 VDMs from the statistical model of Camps \&
Pr\'evot~\shortcite{Camps96} for fluctuations in the geomagnetic
field, using the model parameters proposed for the Icelandic data set.
In this model, the local field vector is the sum of two independent
sets of vectors: a normally distributed axial dipole component plus an
isotropic set of vectors with a Maxwellian distribution that simulates
secular variation.  It is worth pointing out that the trend in the
experimental VDMs for the Brunhes period is very well reproduced in
this statistical model.  We do note that this simulation also predicts
some outliers data -- high VDMs corresponding to excursional VGPs --
as it sometimes observed in the Brunhes experimental data set
(Fig.~\ref{vdm}a) and as has been also recently reported for a mid
Miocene excursion recorded in Canary Island lavas~\cite{Leonhardt00}.

\begin{figure}
\includegraphics{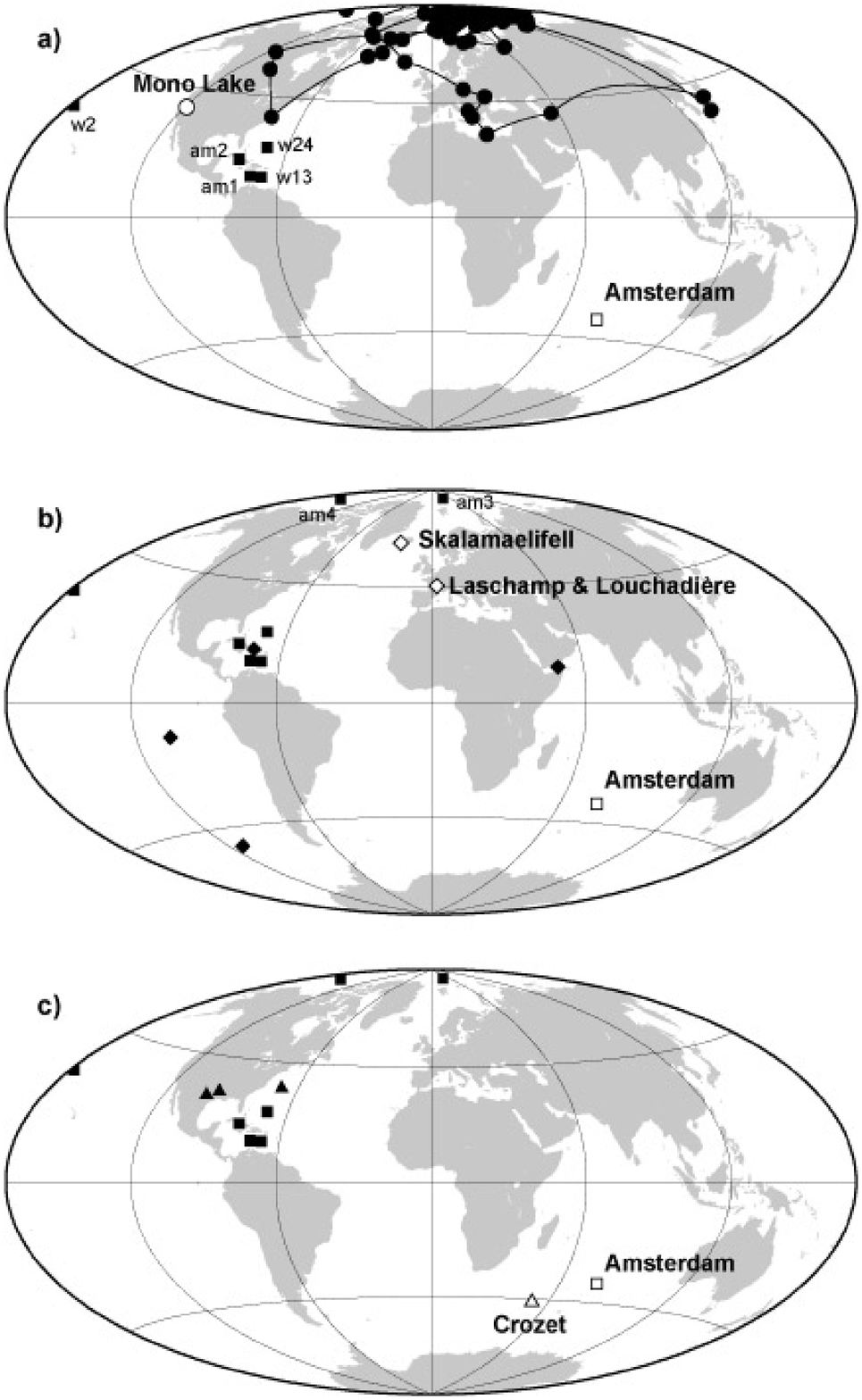}
\caption{Location of the excursional VGPs from Amsterdam Island, black squares, compared to a) VGPs for the Mono Lake record at Wilson Creek (Liddicoat \& Coe, 1979),
  black circles, b) excursional VGP for the Laschamp excursion, black
  diamonds and c) the excursional VGP from Possession Island (Camps \&
  et al., 2001), black triangles. The open symbols represent the
  location of the corresponding sampling sites. We plot Watkins and
  Nougier's (1973) excursional VGPs (w2, w13 and w24) and the present
  study (am1-4) VGPs together.}
\label{mono-vgp}
\end{figure}

Abnormal VGPs from the Amsterdam excursion tend to group over the
Caribbean Sea (Fig.~\ref{mono-vgp}).  The question of whether this
cluster represents a long-lived transitional state of the field or is
an artifact due to a rapid extrusion of successive lavas must be
addressed.  First, Watkins \& Nougier~\shortcite{Watkins73} concluded
from geological field observations that Amsterdam excursion seems to
correspond to two successive departures of the VGPs from the present
geographic pole which recorded very similar VGP locations.  They
argued that one of the excursional flow (their flow 24) belongs to the
old volcanic episode whereas all the other excursional flows (their
flows 13,17-19), although they yield a similar abnormal direction, are
part of a more recent volcanic phase.  We know from experience that
stratigraphic correlations are often speculative in volcanic area,
hence we find their argument quite weak.  However we have no reason to
rule out their conclusion since we did not carry out a field analysis.
Next, we note that one of the excursional VGP from the Laschamp event
(Fig.~\ref{mono-vgp}b) has also a location within the Amsterdam
cluster.  Finally, we point out that Camps et al.~\shortcite{Camps01}
have described a Plio-Pleistocene normal-transitional-normal excursion
recorded in lava flows from Possession Island, a volcanic island
located in the southern India ocean 2300 km southwestward from
Amsterdam.  Interestingly, this excursion is characterized by a
clustering of transitional VGPs also located in the vicinity of the
Caribbean Sea (Fig.~\ref{mono-vgp}c).  Although this excursion is not
radiometrically dated, we are certain from geological considerations
that it is older than the Amsterdam excursion .  Thus, the
transitional field of at least two distinct excursions would have
revisited the same VGP location.  These observations suggest that the
Amsterdam excursional cluster may represents a recurring preferred
location for transitional VGPs like that was previously proposed
\cite{Hoffman92}.

At present, the Ar/Ar isotopic ages suggest that the Amsterdam
excursion corresponds to a late Pleistocene excursion.  During this
period, two geomagnetic events are described in the literature.  It
concerns the Mono Lake excursion, documented in several lacustrine
sedimentary sections from the western USA, which is dated to
$\approx$28 $^{14}$C ka \cite{Liddicoat92}, and the Laschamp
excursion, described for the first time in lava flows from {L}a chaine
des {P}uys, {M}assif {C}entral {F}rance, and dated to $\approx$ 42 ka
(see Kent et al.~\shortcite{Kent02} for a review).  If evidences for
the Laschamp excursion are found in numerous deep-sea sediment
records, this is not the case for the Mono Lake which is only
described jointly with Laschamp from high deposition rate sediments in
the North Atlantic \cite{Nowaczyk97,Laj00}.  At the present time, the
existence of two geomagnetic excursions in the late Pleistocene is
directly questioned by Kent et al.~\shortcite{Kent02}. Using new
$^{14}$C dates on carbonates and \chemical{^{40}Ar/^{39}Ar} sanidine
dates on ash layers, they concluded that Mono Lake excursion at Wilson
Creek should be regarded as a record of the Laschamp excursion.
Unfortunately, the large error bars associated with ours ages estimate
for the Amsterdam excursion do not allow to bring insight into this
debate.  We point out here that our preferred hypothesis, taking into
account the age of 26$\pm$15 and 18$\pm$9 ka obtained for Amsterdam
excursional flows, is to correlate this excursion with a geomagnetic
event younger than 30 Ka which could correspond to Mono Lake if we
consider the former age estimate of 28 Ka \cite{Liddicoat92}.  This
conclusion requires however a further confirmation with a more
accurate age control, but if it is true, Amsterdam excursion would
represent firm evidence for a global occurrence of Mono Lake
excursion.  Finally, it is interesting to point out that the original
record of Mono Lake excursion at Wilson Creek \cite{Liddicoat79}
shows, as the Amsterdam excursion seems to do, two successive
excursional loops (Fig.~\ref{mono-vgp}a).

\section{Conclusions}

\begin{enumerate}
  
\item {We have identified an excursion of the geomagnetic field in the
    late Pleistocene recorded in volcanic rocks from Amsterdam Island,
    Indian Ocean. Good quality alternating field demagnetization
    results show that two flows have excursional polarities with VGP
    latitudes of 15.2\degr and 21.2\degr, and two flows have normal
    polarities (VGP latitudes are 85.5\degr and 77.4\degr).}
  
\item {$^{40}$Ar/$^{39}$Ar dating did not enable us to precisely
    specify which excursion we identified. Mono Lake is the most
    likely, but the large error bars in the final dates do not exclude
    the possibility for theses rocks to have recorded the Laschamp
    excursion. Indeed, some authors believe the two excursions are the
    same \cite{Kent02}}
  
\item{High-quality paleointensity determinations show low VDM values
    for the excursional flows (3.7$\times$10$^{22}$ $\mathrm{Am^2}$
    and 3.4$\times$10$^{22}$ $\mathrm{Am^2}$), whereas normal flows
    have VDMs close to the present-day VDM (6.2$\times$10$^{22}$
    $\mathrm{Am^2}$ and 7.7$\times$10$^{22}$ $\mathrm{Am^2}$). These
    low values are in agreement with other VDMs determinations during
    excursions in the Brunhes period and corroborates the fact that
    the VDM decreases when the colatitude of VGP increases.}
  
\item{For the first time, pTRM-tail-tests from above were used as a
    selection criteria for the paleointensity determinations. We found
    that most of our samples exhibit a tail characteristic of MD
    material for pTRMs given at low temperature and SD-like for
    high-temperature pTRMs. Therefore we rejected measurements
    acquired at low-temperature ($>$ 400\degr C) for the best-fit on
    the Arai plots.}

\end{enumerate}

%
%

\begin{acknowledgments}
  We are grateful to the ``Institut Polaire Paul Emile Victor'' for
  providing all transport facilities and for the support of this
  project.  Special thanks to A. Lamalle and all our field friends. We
  thank M. Pr\'evot, A. Muxworthy, R. Coe , \"{O}. \"{O}zdemir and
  D.J. Dunlop for valuable discussions, A. Goguitchaichvili for
  carrying out the hysteresis measurements, and Anne Delplanque for
  assistance with computer drawing.  The comments of S. Bogue, C.
  Langereis and and anonymous reviewer are appreciated.  This work was
  supported by CNRS-INSU programme int\'erieur Terre.
\end{acknowledgments}

\bibliographystyle{gji}
\bibliography{GC102}

\begin{thebibliography}{52}
\expandafter\ifx\csname natexlab\endcsname\relax\def\natexlab#1{#1}\fi

\bibitem[\protect\citename{Audunsson \& Levi, }1997]{Audunsson97}
Audunsson, H. \& Levi, S., 1997. Geomagnetic fluctuations during a polarity
  transition, {\it J. Geophys. Res.\/}, {\bf 102}(B9), 20259--20268.

\bibitem[\protect\citename{Bol'shakov \& Shcherbakova, }1979]{Bolshakov79}
Bol'shakov, A. \& Shcherbakova, V., 1979. A thermomagnetic criterion for
  determining the domain structure of ferrimagnetics, {\it Izv. Acad. Sci. USSR
  Phys. Solid Earth\/}, {\bf 15}, 111--117.

\bibitem[\protect\citename{Calvo et~al., }2002]{Calvo02}
Calvo, M., Pr\'evot, M., Perrin, M., \& Riisager, J., 2002. Investigating the
  reasons for the failure of paleointensity experiments: A study on historical
  lava flows from mt. etna (italy), {\it Geophys. J. Int.\/}, p. Under Press.

\bibitem[\protect\citename{Camps \& Pr\'evot, }1996]{Camps96}
Camps, P. \& Pr\'evot, M., 1996. A statistical model of the fluctuations in the
  geomagnetic field from paleosecular variation to reversal, {\it Sciences\/},
  {\bf 273}, 776--779.

\bibitem[\protect\citename{Camps et~al., }2001]{Camps01}
Camps, P., Henry, B., Pr\'{e}vot, M., \& Faynot, L., 2001. Geomagnetic
  paleosecular variation recorded in plio-pleistocene volcanic rocks from
  {P}ossession {I}sland ({C}rozet {A}rchipelago, southern {I}ndian {O}cean,
  {\it J. Geophys. Res.\/}, {\bf 106}(B2), 1961--1972.

\bibitem[\protect\citename{Carvallo \& Dunlop, }2001]{Carvallo01}
Carvallo, C. \& Dunlop, D., 2001. Archeomagnetism of potsherds from {G}rand
  {B}anks, {O}ntario: a test of low paleointensities in {O}ntario around
  {A}.{D}.1000., {\it Earth Planet. Sci. Lett.\/}, {\bf 186}, 437--450.

\bibitem[\protect\citename{Chauvin et~al., }1989]{Chauvin89}
Chauvin, A., Duncan, R., Bonhommet, N., \& Levi, S., 1989. Paleointensity of
  the {E}arth's magnetic field and {K}-{A}r dating of the {L}ouchadi\`ere flow
  ({C}entral {F}rance), new evidence for the {L}aschamps excursion, {\it
  Geophys. Res. Lett.\/}, {\bf 16}, 1189--1192.

\bibitem[\protect\citename{Coe et~al., }1978]{Coe78}
Coe, R., Gromm\'{e}, C., \& Mankinen, E., 1978. Geomagnetic paleointensities
  from radiocarbon-dated lava flows on {H}awaii and the question of the
  {P}acific nondipole low, {\it J. Geophys. Res.\/}, {\bf 83}, 1740--1756.

\bibitem[\protect\citename{Day et~al., }1977]{Day77}
Day, M., Fuller, M., \& Schmidt, V., 1977. Hysteresis properties of
  titanomagnetites: grain size and compositional dependance., {\it Phys. Earth
  Planet. Int\/}, {\bf 13}, 267--267.

\bibitem[\protect\citename{Deckart et~al., }1997]{Deckart97}
Deckart, K., F\'eraud, G., \& Bertrand, H., 1997. Age of {J}urassic continental
  tholeiites of {F}rench {G}uyana, {S}urinam and {G}uinea; implications for the
  initial opening of the central {A}tlantic {O}cean, {\it Earth Planet. Sci.
  Lett.\/}, {\bf 150}(3-4), 205--220.

\bibitem[\protect\citename{Dunlop \& \"{O}zdemir, }1997]{Dunlop97}
Dunlop, D. \& \"{O}zdemir, O., 1997. {\it Rock magnetism: {F}undamentals and
  frontiers\/}, Cambridge Univ. Press, New-York, 573 pp.

\bibitem[\protect\citename{Dunlop \& \"{O}zdemir, }2001]{Dunlop01}
Dunlop, D. \& \"{O}zdemir, O., 2001. Beyond {N}\'eel's theories: thermal
  demanetization of narrow-band partial thermoremanent magnetizations., {\it
  Phys. Earth Planet. int.\/}, {\bf 126}(1-2), 43--57.

\bibitem[\protect\citename{Graham et~al., }1999]{Graham99}
Graham, D., Johnson, K., Douglas~Priebe, L., \& Lupton, J., 1999. Hotspot-ridge
  interaction along the {S}outheast {I}ndian {R}idge near {A}msterdam and {S}t
  {P}aul islands: helium isotope evidence, {\it Earth Planet. Sci. Lett.\/},
  {\bf 167}, 297--310.

\bibitem[\protect\citename{Gubbins, }1999]{Gubbins99}
Gubbins, D., 1999. The distinction between geomagnetic excursions and
  reversals, {\it Geophys. J. Int.\/}, {\bf 137}, F1--F3.

\bibitem[\protect\citename{Gunn et~al., }1971]{Gunn71}
Gunn, B., Abranson, C., Nougier, J., Watkins, N., \& Hajash, A., 1971.
  Amsterdam island, an isolated volcano in the {S}outhern {I}ndian {O}cean,
  {\it Contr. Mineral. and Petrol.\/}, {\bf 32}, 79--92.

\bibitem[\protect\citename{Guyodo \& Valet, }1999]{Guyodo99}
Guyodo, Y. \& Valet, J., 1999. Global changes in intensity of the {E}arth's
  magnetic field during the past 800 kyr, {\it Nature\/}, {\bf 399}, 249--252.

\bibitem[\protect\citename{Hanes et~al., }1985]{Hanes85}
Hanes, J., York, D., \& Hall, C., 1985. An $\mathrm{^{40}{A}r/^{39}{A}r}$
  geochronological and electron microprobe investigation of an {A}rchean
  pyroxenite and its bearing on ancient atmospheric compositions, {\it Can. J.
  Earth Sci.\/}, {\bf 22}, 947--958.

\bibitem[\protect\citename{Hoffman, }1981]{Hoffman81}
Hoffman, K., 1981. Palaeomagnetic excursions, aborted reversals and
  transitional fields, {\it Nature\/}, {\bf 294}, 67--69.

\bibitem[\protect\citename{Hoffman, }1992]{Hoffman92}
Hoffman, K., 1992. Dipolar reversal states of the geomagnetic field and
  core-mantle dynamics, {\it Nature\/}, {\bf 359}, 789--794.

\bibitem[\protect\citename{Johnson et~al., }2000]{Johnson00}
Johnson, K., Graham, D., Rubin, K., Nicolaysen, K., Scheirer, D., Forsyth,
  D.~W., Baker, E.~T., \& Douglas-Priebe, L.~M., 2000. Boomerang {S}eamount;
  the active expression of the {A}msterdam-{S}t. {P}aul {H}otspot, {S}outheast
  {I}ndian {R}idge, {\it Earth and Planet. Sci. Lett.\/}, {\bf 183}, 245--259.

\bibitem[\protect\citename{Kent et~al., }2002]{Kent02}
Kent, D., Hemming, S., \& Turrin, B., 2002. Laschamp excursion at {M}ono
  {L}ake?, {\it Earth Planet. Sci. Lett.\/}, {\bf 197}, 151--164.

\bibitem[\protect\citename{Kirschvink, }1980]{Kirschvink80}
Kirschvink, J., 1980. The least-squares line and plane and the analysis of
  paleomagnetic data, {\it Geophys. J. R. astr. Soc.\/}, {\bf 62}, 699--718.

\bibitem[\protect\citename{Laj et~al., }2000]{Laj00}
Laj, C., Kissel, C., Mazaud, A., Channell, J., \& Beer, J., 2000. North
  {A}tlantic palaeointensity stack since 75 ka (napis-75) and the duration of
  the {L}aschamp event, {\it Phil. Trans. R. Soc. Lond. A\/}, {\bf 358},
  1009--1025.

\bibitem[\protect\citename{Langereis et~al., }1997]{Langereis97}
Langereis, C., Dekkers, M., de~Lange, G., Paterne, M., \& van Santvoort, P.,
  1997. Magnetostratigraphy and astronomical calibration of the last 1.1 {M}yr
  from an eastern {M}editerranean piston core and dating of short events in the
  {B}ruhnes, {\it Geophys. J. Int.\/}, {\bf 129}, 75--94.

\bibitem[\protect\citename{Leonhardt et~al., }2000]{Leonhardt00}
Leonhardt, R., Hufenbecher, F., Heider, F., \& Soffel, H., 2000. High absolute
  paleointensity during a mid {M}iocene excursion of the {E}arth's magnetic
  field, {\it Earth Planet. Sci. Lett.\/}, {\bf 184}, 141--154.

\bibitem[\protect\citename{Levi \& Karlin, }1989]{Levi89}
Levi, S. \& Karlin, R., 1989. A sixty thousand year paleomagnetic record from
  {G}ulf of {C}alifornia sediments: secular variation, late {Q}uaternary
  excursions and geomagnetic implications, {\it Earth Planet. Sci. Lett.\/},
  {\bf 92}, 219--233.

\bibitem[\protect\citename{Levi et~al., }1990]{Levi90}
Levi, S., Audunaaon, H., Duncan, R., Kristjansson, L., Gillot, P., \&
  Jacobsson, S., 1990. Late {P}leistocene geomagnetic excursion in {I}celandic
  lavas: confirmation of the {L}aschamps excursion, {\it Earth Planet. Sci.
  Lett.\/}, {\bf 96}, 443--457.

\bibitem[\protect\citename{Liddicoat, }1992]{Liddicoat92}
Liddicoat, J., 1992. Mono {L}ake excursion in {M}ono {B}asin, {C}alifornia, and
  at {C}arson {S}ink and {P}yramid {L}ake, {N}evada, {\it Geophys. J. Int.\/},
  {\bf 108}, 442--452.

\bibitem[\protect\citename{Liddicoat, }1996]{Liddicoat96}
Liddicoat, J., 1996. Mono {L}ake excursion in the {L}ahontan {B}asin, {N}evada,
  {\it Geophys. J. Int.\/}, {\bf 125}, 630--635.

\bibitem[\protect\citename{Liddicoat \& Coe, }1979]{Liddicoat79}
Liddicoat, J. \& Coe, R., 1979. Mono {L}ake geomagnetic excursion, {\it J.
  Geophys. Res\/}, {\bf 1984}, 261--271.

\bibitem[\protect\citename{Nagata et~al., }1963]{Nagata63}
Nagata, T., Arai, Y., \& Momose, K., 1963. Secular variation of the geomagnetic
  total force during the last 5,000 years, {\it J. Geophys. Res.\/}, {\bf 68},
  5277--5282.

\bibitem[\protect\citename{Nowaczyk \& Antonow, }1997]{Nowaczyk97}
Nowaczyk, N. \& Antonow, M., 1997. High-resolution magnetostratigraphy of four
  sediment cores from the {G}reenland {S}ea -- {I}. {I}dentification of the
  {M}ono {L}ake excursion, {L}aschamp and {B}iwa {I}/{J}amaica geomagnetic
  polarity events, {\it Geophys. J. Int.\/}, {\bf 131}, 310--324.

\bibitem[\protect\citename{Perrin, }1998]{Perrin98}
Perrin, M., 1998. Paleointensity determination, magnetic domain structure, and
  selection criteria, {\it J. Geophys. Res.\/}, {\bf 103}, 30591--30600.

\bibitem[\protect\citename{Pr\'evot et~al., }1983]{Prevot83}
Pr\'evot, M., Mankinen, E., Gromm\'e, C., \& Lecaille, A., 1983. High
  paleointensities of the geomagnetic field from thermomagnetic study on rift
  valley pillow basalts from {M}id-{A}tlantic {R}idge, {\it J. Geophys.
  Res.\/}, {\bf 88}, 2316--2326.

\bibitem[\protect\citename{Quidelleur et~al., }1999]{Quidelleur99}
Quidelleur, X., Gillot, P., Carlut, J., \& Courtillot, V., 1999. Link between
  excursions and paleointensity inferred from abnormal field directions
  recorded at {L}a {P}alma around 600 ka, {\it Earth Planet. Sci. Lett.\/},
  {\bf 168}, 233--242.

\bibitem[\protect\citename{Renne et~al., }1998]{Renne98}
Renne, P., Swisher, C., Deino, A., Karner, D., Owens, T., \& DePaolo, D., 1998.
  Intercalibration of standards, absolute ages and uncertainties in $\mathrm{
  ^{40}{A}r / ^{39}{A}r}$ dating, {\it Chem. Geol.\/}, {\bf 154}, 117--152.

\bibitem[\protect\citename{Roddick et~al., }1980]{Roddick80}
Roddick, J., Cliff, R., \& Rex, D., 1980. The evolution of excess argon in
  alpine biotites- {A} $\mathrm{^{40}{A}r/^{39}{A}r}$ analysis, {\it Earth
  Planet Sci. Lett\/}, {\bf 48}, 185--208.

\bibitem[\protect\citename{Roperch et~al., }1988]{Roperch88}
Roperch, P., Bonhommet, N., \& Levi, S., 1988. Paleointensity of the {E}arth's
  magnetic field during the {L}aschamps excursion and its geomagnetic
  implications, {\it Earth Planet. Sci. Lett.\/}, {\bf 88}, 209--219.

\bibitem[\protect\citename{Royer \& Schlich, }1988]{Royer88}
Royer, J. \& Schlich, R., 1988. {S}outheast {I}ndian {R}idge between the
  {R}odriguez triple junction and the {A}msterdam and {S}aint {P}aul {I}slands:
  detailed kinematics for the past 20 m.y., {\it J. Geophys. Res.\/}, {\bf 93},
  13524--13550.

\bibitem[\protect\citename{Ruffet et~al., }1991]{Ruffet91}
Ruffet, G., F\'{e}raud, G., \& Amouric, M., 1991. Comparison of $\mathrm{
  ^{40}{A}r / ^{39}Ar}$ conventional and laser dating biotites from the {N}orth
  {T}r\'egor {B}atholith, {\it Geochim. et Cosmochim. Acta\/}, {\bf 55},
  1675--1688.

\bibitem[\protect\citename{Schnepp \& Hradetzky, }1994]{Schnepp94}
Schnepp, E. \& Hradetzky, H., 1994. Combined paleointensity and
  $\mathrm{^{40}{A}r/^{39}{A}r}$ age spectrum data from volcanic rocks of the
  {W}est {E}ifel field ({G}ermany): evidence for an early {B}runhes geomagnetic
  excursion, {\it J. Geophys. Res.\/}, {\bf 99}(B5), 9061--9076.

\bibitem[\protect\citename{Selkin \& Tauxe, }2000]{Selkin00}
Selkin, P. \& Tauxe, L., 2000. Long term variations in palaeointensity, {\it
  Phil. Mag.\/}, {\bf 358}(1768), 1065--1088.

\bibitem[\protect\citename{Shcherbakov \& Shcherbakova, }2001]{Shcherbakov01}
Shcherbakov, V. \& Shcherbakova, V., 2001. On the suitability of the {T}hellier
  method of paleointensity determinations on pseudo-single-domain and
  multidomain grains., {\it Geophys. J. Int.\/}, {\bf 146}, 20--30.

\bibitem[\protect\citename{Shcherbakova et~al., }2000]{Shcherbakova00}
Shcherbakova, V., Shcherbakov, V., \& Heider, F., 2000. Properties of partial
  thermoremanent magnetization in pseudosingle domain and multidomain magnetite
  grains, {\it J. Geophys. Res.\/}, {\bf 105}, 77767--781.

\bibitem[\protect\citename{Small, }1995]{Small95}
Small, C., 1995. Observation of ridge-hotspot interactions in the {S}outhern
  {O}cean, {\it J. Geophys. Res.\/}, {\bf 100}(B9), 17931--17946.

\bibitem[\protect\citename{Tauxe, }1998]{Tauxe98}
Tauxe, L., 1998. {\it Paleomagnetic {P}rinciples and {P}ractice\/}, Kluwer,
  Dordrecht.

\bibitem[\protect\citename{Thellier \& Thellier, }1944]{Thellier44}
Thellier, E. \& Thellier, O., 1944. Recherches g\'eomagn\'etiques sur des
  coul\'ees volcaniques d'{A}uvergne, {\it Ann. Geophys.\/}, {\bf 1}, 37--52.

\bibitem[\protect\citename{Thellier \& Thellier, }1959]{Thellier59}
Thellier, E. \& Thellier, O., 1959. Sur l'intensit\'e du champ magn\'etique
  terrestre dans le pass\'e historique et g\'eologique, {\it Ann. Geophys.\/},
  {\bf 15}, 285--376.

\bibitem[\protect\citename{Turner, }1971]{Turner71}
Turner, G., 1971. $\mathrm{^{40}Ar/^{39}Ar}$ ages from the lunar {M}aria, {\it
  Earth Planet. Sci. Lett.\/}, {\bf 11}, 169--191.

\bibitem[\protect\citename{Vinogradov \& Markov, }1989]{Vinogradov89}
Vinogradov, Y. \& Markov, G., 1989. {\it On the effect of low temperature
  heating on the magnetic state of multi-domain magnetite- {I}nvestigations in
  {R}ockmagnetism and {P}aleomagnetism\/}, pp. 31--39, Institute of Physics of
  the Earth, Moscow.

\bibitem[\protect\citename{Watkins \& Nougier, }1973]{Watkins73}
Watkins, N. \& Nougier, J., 1973. Excursions and secular variations of the
  {B}runhes epoch geomagnetic field in the {I}ndian ocean region, {\it J.
  Geophys. Res.\/}, {\bf 78}(26), 6060--6068.

\bibitem[\protect\citename{Zhu et~al., }2000]{Zhu00}
Zhu, R., Pan, Y., \& Coe, R., 2000. Paleointensity studies of a lava succession
  from {J}ilin {P}rovince, northeastern {C}hina: {E}vidence for the {B}lake
  event, {\it J. Geophys. Res.\/}, {\bf 105}, 8305--8317.

\end{thebibliography}

\newpage \clearpage

\newpage

\clearpage

%
%

\begin{table*}
\begin{minipage}{130mm}
\caption{Directional Results}
\label{directions}
\begin{tabular}{cccrrrrrrrr}
Flow& n/N& Inc.& Dec.& $\alpha_{95}$& $\kappa$& $\delta$& Lat.& Long.& $\chi$&  $\mathrm{J_{20}}$ \\
\hline

am1 &7/7 &-70.7 &238.6 &5.0 &144  &45.6  &15.2 &287.8 &1721.2 &3.51 \\
am2 &7/7 &-76.2 &238.2 &1.3 &2287 &41.6  &21.2 &281.1 &1812.9 &2.91 \\
am3 &7/8 &-52.5 &356.1 &4.1 &217  &5.2   &85.5 &40.5  &1740.4 &7.95 \\
am4 &7/7 &-63.1 &3.9   &3.8 &249  &9.1   &77.4 &205.6 &661.1  &6.49 \\
\hline
\end{tabular}

\medskip
n/N is the number of samples used in the analysis/total number of samples collected;
Inc. and Dec. are the mean inclination positive downward and declination east of north, 
respectively;
$\alpha_{95}$ and $\kappa$ are the  95\% confidence cone about average direction and the concentration parameter of Fisher statistics, respectively;
$\delta$ is the reversal angle measured in degrees from the direction of the dipole axial field direction;
Lat. and Long. are latitude and longitude of corresponding VGP position, respectively;
$\chi$ is the geometric mean susceptibility ($\times 10^{-6}$ SI);
$\mathrm{J_{20}A/m}$ is the geometric mean remanence intensity in A/m measured after a magnetic treatment of 20 mT. The flow am1 corresponds to Watkins and Nougier's (1973) flows 17 and 18 combined (see text for explanation) and am2 to their flow 19.   

\end{minipage}
\end{table*}

\begin{table*}
\begin{minipage}{150mm}
\caption{Thermomagnetic and rock magnetism properties of thermally stable samples}
\label{a-parameter}
\begin{tabular}{lllllllll}
Sample& T$_C$ &MDF&M$_{rs}$/M$_s$ &H$_{cr}$/H$_c$ &{300-Tr}
&{400-300}&{500-400}&{550-500}\\ 
      &       &   &        &               &A (B)        &A (B)         &A (B)         &A (B)\\
\hline
\multicolumn{9}{l}{{\it Flow am1}}\\
668 &481$\pm$21, 581$\pm$14      &47 &0.10 &3.95  &31 (13) &23 (10) &10 (33) &4 (44)\\
669 &480$\pm$32, 540$\pm$10      &30 &0.06 &3.72  &43 (10) &60 (~7) &29 (30) &8 (53)\\
670 &394$\pm$88, 572$\pm$8       &53 &0.17 &3.31  &36 (17) &35 (14) &9  (35) &5 (34)\\
671 &485$\pm$15, 587$\pm$20      &38 &0.17 &8.13  &29 (13) &32 (~9) &12 (36) &4 (42)\\
673 &503$\pm$38                  &34 &n.d  &n.d   &27 (12) &30 (10) &7  (51) &6 (27)\\
674 &475$\pm$11, 572$\pm$18      &56 &0.22 &2.54  &26 (nd) &26 (nd) &8  (nd) &nd  (nd)\\
\multicolumn{9}{l}{{\it Flow am2}}\\
675 &473$\pm$68                  &27 &0.25 &1.76  &32 (~7) &37 (~6) &11 (30) &2 (57)\\
680 &516$\pm$54                  &23 &0.19 &1.83  &46 (14) &60 (13) &11 (42) &3 (31)\\
\multicolumn{9}{l}{{\it Flow am3}}\\
690 &564$\pm$20                  &53 &0.24 &2.13  &35 (~8) &40 (~7) &11 (30) &4 (55)\\
691 &504$\pm$30, 561$\pm$17      &56 &0.25 &1.96  &38 (~7) &43 (~6) &12 (30) &4 (57)\\
692 &500$\pm$9,  560$\pm$20      &49 &0.25 &1.98  &33 (~7) &37 (~5) &13 (27) &2 (61)\\
694 &552$\pm$23                  &41 &0.23 &1.89  &42 (~5) &38 (~5) &10 (24) &2 (66)\\
695 &510$\pm$32, 561$\pm$19      &48 &n.d  &n.d   &46 (~7) &53 (~6) &10 (32) &3 (55)\\
\multicolumn{9}{l}{{\it Flow am4}}\\
684 &569$\pm$23                  &39 &0.32 &1.62  &44 (~4) &37 (~5) &8  (23) &3 (68)\\
685 &505$\pm$34, 547$\pm$7       &30 &n.d  &n.d   &49 (10) &42 (10) &8  (40) &5 (40)\\
686 &550$\pm$10                  &33 &n.d  &n.d   &48 (22) &40 (21) &8  (32) &6 (25)  \\
\hline
\end{tabular}
\medskip
$\mathrm{T_{C}}$ is the mean Curie temperature (\degr C) calculated according to the method of Pr\'evot
et al.~\shortcite{Prevot83}. The confidence intervals for the Curie temperatures indicate the temperature 
range in which the KT curve correspond to a straight line. 
MDF is the Median Destructive alternating Field in mT; $\mathrm{M_{rs}/M_{s}}$
and $\mathrm{H_{cr}/H_{c}}$ are the hysteresis parameters; A values are the 
relative intensities measured at room temperature of the pTRM tail expressed in percent
$\mathrm{A(T_1,T_2) = tail[pTRM(T_1,T_2)]/pTRM(T_1,T_2)}$. 
B values shown in parentheses correspond to the percent of the total pTRM 
(e.g., $\mathrm{\sum_{i}pTRM_i}$) each pTRM(T$_1$,T$_2$) represents; 
intensities are measured at room temperature.  
nd means not determined. Note that sample 674 broke itself before the acquisition of the pTRM(550-500) 
was completed. Thus, we were not able to calculate B values for this sample. 
\end{minipage}
\end{table*}

\begin{table*}
\begin{minipage}{170mm}
\caption{Accepted paleointensity determinations.}
\label{table-pi}
\begin{tabular}{llccccccccccc}
Flow &Sample &Fe\( \pm \sigma\) Fe &$\mathrm{T_{1}-T_{2}}$ &n &f &g &q &MAD &$\alpha$ &Drat &$\bar{F}_E$\( \pm\) s.d. &VDM\\
\hline
am1 &670D &19.5\( \pm \)0.6  &400-570 &8 &0.643 &0.845 &16.9 &6.5 &2.6 &5.1 &24.6\( \pm \)3.7 &3.7 \\
    &671D &27.1\( \pm \)0.7  &400-580 &9 &0.688 &0.867 &23.1 &3.8 &4.7 &0.9 & & \\
    &673C &27.6\( \pm \)0.5  &400-570 &8 &0.844 &0.806 &36.8 &2.7 &1.3 &0.9 & &  \\
    &674C &24.3\( \pm \)0.6  &400-570 &8 &0.754 &0.839 &27.7 &3.7 &1.6 &0.8 & &  \\

am2 &675E &22.1\( \pm \)0.7  &400-550 &6 &0.862 &0.729 &19.8 &3.3 &2.3 &1.8  &24.0\( \pm \)2.6 &3.4\\
    &680C &25.8\( \pm \)1.8  &400-550 &6 &0.435 &0.757 &4.8  &2.7 &2.2 &10.0 & &\\

am3 &690B &30.3\( \pm \)0.6  &400-580 &9 &0.892 &0.848 &35.8 &3.2 &1.2 &2.3  &32.8\( \pm \)2.2 &6.2\\ 
    &691C &34.9\( \pm \)0.4  &400-580 &8 &0.925 &0.755 &63.2 &3.4 &1.3 &2.3  & &\\
    &692D &33.3\( \pm \)0.9  &400-570 &8 &0.939 &0.797 &28.2 &3.3 &1.2 &2.6  & &\\
    &694D &34.9\( \pm \)0.8  &400-570 &8 &0.935 &0.798 &34.1 &4.3 &1.1 &2.5  & &\\
    &695B &30.7\( \pm \)0.6  &400-570 &8 &0.927 &0.821 &39.4 &4.6 &0.8 &3.0  & &\\

am4 &684E &42.3\( \pm \)1.4  &400-550 &6 &0.700 &0.777 &16.7 &3.3 &2.1 &2.2  &46.9\( \pm \)4.6 &7.7\\
    &685E &47.0\( \pm \)0.8  &400-570 &7 &0.758 &0.742 &34.4 &1.7 &1.5 &2.6  & &\\
    &686E &51.4\( \pm \)1.1  &400-570 &8 &0.790 &0.814 &31.3 &3.4 &2.3 &2.5  & &\\
\hline
\end{tabular}

\medskip
Sample is an identifier of a sample used for the paleointensity determination;
Fe is paleointensity estimate (in $\mu T$) for individual specimen, and (\(\sigma\) Fe) is its
standard error;
$\mathrm{T_{1} and T_{2}}$ are the minimum and maximum of the temperature range in \degr C used to determine
paleointensity;
n is the number of points in the $\mathrm{T_{1}-T_{2}}$ interval;
f, g, and q are NRM fraction, gap factor and quality factor, respectively \cite{Coe78};
MAD is the maximum angular deviation calculated along with the principal component for the NRM left in the
$\mathrm{T_{1}-T_{2}}$ interval;
$\alpha$ is the angle in degrees between the vector average and the principal component calculated for 
the NRM left in the $\mathrm{T_{1}-T_{2}}$ interval;
Drat is expressed in percent and corresponds to the difference ratio between repeat pTRM steps normalized
by the length of the selected NRM-pTRM segment \cite{Selkin00};
$\bar{F}_E$ is unweighted average paleointensity of individual lava flow and its standard deviation and
VDM is the corresponding virtual dipole moment ($\times 10^{22} \mathrm{Am^2}$).

\end{minipage}
\end{table*}

\begin{table*}
\begin{minipage}{130mm}
\caption{Isotopic ages Results}
\label{dating}
\begin{tabular}{ccccc}
Flow& Sample& $\mathrm{(^{40}Ar/^{39}Ar)_i \pm 2\sigma}$ &Isochrone age
(ka$\pm$ka) &MSWD\\ \hline
am1 &674B &290.4$\pm$0.8 &26$\pm$15 &1.5 \\
am2 &676D &288.6$\pm$1.5 &18$\pm$9  &0.2 \\
am3 &694F &290.2$\pm$1.2 &48$\pm$22 &3.2 \\
\hline
\end{tabular}
\medskip
\end{minipage}
\end{table*}

\begin{table*}
\begin{minipage}{80mm}
\caption{Laboratory Koenigsberger ratios before and after heating.}
\label{koenig}
\begin{tabular}{llcc}
Flow &Sample &$Q_L$ &$Q'_L$ \\ \hline
am1 &670D &14.4 & 21.7 \\
    &671D &17.9 & 20.4 \\
    &673C &28.0 & 27.8 \\
    &674C &28.6 & 27.9 \\

am2 &675E &39.5 & 34.5 \\
    &680C &8.1  & 8.3  \\
 
am3 &690B &32.2 & 29.3 \\
    &691C &36.1 & 35.4 \\
    &692D &34.8 & 33.7 \\
    &694D &32.2 & 31.7 \\
    &695B &30.2 & 29.6 \\

am4 &684E &42.1 & 41.7 \\
    &685E &37.3 & 29.7 \\
    &686E &33.9 & 27.4 \\
\hline
\end{tabular}

\medskip
Sample is the same identifier than used for the paleointensity determination;
$Q_L$ is the laboratory Koenigsberger ratio before heating; $Q'_L$ is the Koenigsberger ratio after
heating. 
\end{minipage}
\end{table*}


\label{lastpage}

\end{document}